\documentclass[preprint,showpacs,preprintnumbers,amsmath,amssymb]{revtex4}

\usepackage{graphicx,color}

\begin{document}

\title{
The effects of finite distance on the gravitational 
deflection angle of light
} 
\author{Toshiaki Ono}
\author{Hideki Asada} 
%\email{asada@phys.hirosaki-u.ac.jp}
\affiliation{
Graduate School of Science and Technology, Hirosaki University,
Aomori 036-8561, Japan} 
\date{\today}

%\abstract{
\begin{abstract}
In order to clarify effects of 
the finite distance from a lens object to 
a light source and a receiver, the gravitational  deflection of light has been recently reexamined by  using the Gauss-Bonnet (GB) theorem in differential geometry  
[Ishihara et al. 2016]. 
The purpose of the present paper is to give a short review 
of a series of works initiated by the above paper. 
First, we provide the definition of the gravitational deflection angle 
of light for the finite-distance source and receiver 
in a static, spherically symmetric and asymptotically flat spacetime. 
We discuss the geometrical invariance of the definition 
by using the GB theorem. 
The present definition is used to discuss finite-distance effects 
on the light deflection in Schwarzschild spacetime, 
for both cases of the weak deflection and 
strong deflection. 
Next, we extend the definition to stationary and axisymmetric 
spacetimes. 
We compute finite-distance effects on 
the deflection angle of light for Kerr black holes and rotating Teo wormholes. 
Our results are consistent with the previous works 
if we take the infinite-distance limit.
We briefly mention also the finite-distance effects on the light deflection 
by Sagittarius A$^*$. 
%}
\end{abstract}

\maketitle

\section{Introduction}
In 1919, the experimental confirmation of the theory of general relativity 
\cite{GR} succeeded \cite{Eddington}. 
It is the measurement of the gravitational deflection angle of light. 
Since then, the gravitational deflection angle of light has attracted 
a lot of attention. 
Many authors have studied the gravitational deflection of light by black holes 
\cite{Hagihara, Ch, MTW, Darwin, Bozza, Iyer, Bozza+, Frittelli, VE2000, Virbhadra, VNC, VE2002, VK2008, Zschocke}. 
The gravitational lens by other objects such as wormholes and gravitational monopoles 
also has attracted a lot of interest  
\cite{ERT, Perlick, Abe, Toki, Nakajima, Gibbons, DA, Kitamura,Tsukamoto,Izumi,Kitamura2014,Nakajima2014,Tsukamoto2014,Azreg}. 
Very recently, the EHT team has reported a direct image 
of the inner edge of the hot matter around the black hole candidate 
at the center of M87 galaxy \cite{EHT}. 
The direct imaging of black hole shadows must again and steeply raise 
the importance of the gravitational deflection of light.

Most of those calculations are based on the coordinate angle. The angle respects the rotational symmetry of the spacetime. 
Gibbons and Werner (2008) made an attempt of defining, 
in a more geometrical manner, 
the deflection angle of light \cite{GW2008}. 
In their paper, the source and receiver are needed to be located 
at an asymptotic Minkowskian region. 
The Gauss-Bonnet theorem was applied 
to a spatial domain with introducing 
the optical metric, 
for which a light ray is expressed 
as a spatial geodesic  curve. 
Ishihara et al. have successfully extended Gibbons and Werner's idea, 
such that the source and receiver can be at a finite distance 
from the lens object 
\cite{Ishihara2016}. 
They extend the earlier work to 
the case of the strong deflection limit, 
in which the winding number of the photon orbits may be 
larger than unity \cite{Ishihara2017}. 
In particular, the asymptotic receiver and source are not needed. 
Arakida \cite{Arakida2018} made an attempt to apply the Gauss-Bonnet theorem to quadrilaterals that are not extending to infinity and proposed a new definition of 
the deflection angle of light, though a comparison between two different manifolds that he proposed is an open issue. 
With proposing an alternative definition of the  deflection angle of light, Crisnejo et al. \cite{Crisnejo2019} has recently made a  comparison between the alternative definitions in \cite{Ishihara2016,Ishihara2017,Arakida2018} 
and shown by explicit calculations that the definition 
by Arakida in \cite{Arakida2018} is different from 
that by Ishihara et al. \cite{Ishihara2016,Ishihara2017}. 
Their definition has been applied to study the gravitational lensing with a plasma medium \cite{Crisnejo2019}.

The earlier works \cite{Ishihara2016, Ishihara2017} 
are restricted within the spherical symmetry. 
Ono et al. have extended the Gauss-Bonnet method with the optical metric 
to axisymmetric spacetimes \cite{Ono2017}. 
This extension includes mathematical quantities and calculations, 
with which most of the physicists are not very familiar. 
Therefore, the purpose of this paper provides a review of 
the series of papers on the gravitational deflection of light 
for finite-distance source and receiver. 
In particular, we hope that 
detailed calculations in this paper will 
be helpful for readers to compute the gravitational deflection 
of light by the new powerful method. 
For instance, this new technique has been used to 
study the gravitational lensing in rotating Teo wormholes \cite{Ono2018} 
and also in Damour-Solodukhin wormholes \cite{Ovgun2018}. 
This formulation has been successfully used to 
clarify the deflection of light in a rotating global 
monopole spacetime with a deficit angle \cite{Ono2019}.

This paper is organized as follows. 
Section II discusses the definition of the gravitational deflection angle 
of light in static and spherically symmetric spacetimes. 
Section III considers the weak deflection of light 
in Schwarzschild spacetime. 
Section IV discusses the weak deflection of light 
in the Kottler spacetime and the Weyl conformal gravity model. 
The strong deflection of light is examined in Section V. 
Sagittarius A$^*$ (Sgr A$^*$) is also discussed 
as an example for possible candidates. 
In section VI, we discuss the strong deflection of light 
with finite-distance corrections 
in Schwarzschild spacetime. 
Section VII proposes the definition of 
the gravitational deflection angle of light 
in stationary and axisymmetric spacetimes. 
Sgr A$^*$ is also discussed. 
The weak deflection of light is discussed 
for Kerr spacetime in Section VIII 
and 
for rotating Teo wormholes in Section IX. 
Section X is the summary of this paper. 
Appendix A provides the detailed calculations for the Kerr spacetime. 
Throughout this paper, we use the unit of $G=c=1$, 
and the observer may be called the receiver 
in order to avoid confusion between $r_O$ and $r_0$ by using $r_R$.

\section{Definition of the gravitational deflection  angle of light: 
Static and spherically symmetric spacetimes} 
\subsection{Notation}
Following Ishihara et al. \cite{Ishihara2016}, 
this section begins with considering 
a static and spherically symmetric (SSS) spacetime. 
The metric of this spacetime can be written as 
\begin{eqnarray}
ds^2 &=& g_{\mu\nu} dx^{\mu} dx^{\nu} 
\nonumber\\
&=& -A(r) dt^2 + B(r) dr^2 + r^2 d\Omega^2 , 
\label{ds2-SSS-AB}
\end{eqnarray}
where 
$d\Omega^2 \equiv d\theta^2 + \sin^2\theta d\phi^2$, 
and $t$, $\theta$ and $\phi$ are associated with 
the symmetries of the SSS spacetime. 
For a metric of the form (1) we always have to restrict to the domain where $A(r)$ and $B(r)$ are positive, 
such that a static emitter and a static receiver can exist.
The spacetime has a spherical symmetry. 
Therefore, 
the photon orbital plane is chosen, 
without loss of generality, 
as the equatorial plane 
($\theta = \pi/2$). 
We follow the usual definition of 
the impact parameter of the light ray as 
\begin{eqnarray}
b &\equiv& \frac{L}{E} 
\nonumber\\
&=& \frac{r^2}{A(r)} \frac{d\phi}{dt} . 
\label{b}
\end{eqnarray}
From $ds^2=0$ for the light ray, 
the orbit equation is derived as 
\begin{eqnarray}
\left( \frac{dr}{d\phi} \right)^2 
+ \frac{r^2}{B(r)} 
= \frac{r^4}{b^2 A(r)B(r)} . 
\label{orbiteq}
\end{eqnarray}

%\subsection{Optical metric}
Light rays are described 
by the null condition $ds^2 = 0$, 
which is solved for $dt^2$ as 
\begin{eqnarray}
dt^2 &=& \gamma_{IJ} dx^I dx^J 
\nonumber\\
&=& \frac{B(r)}{A(r)} dr^2 + \frac{r^2}{A(r)} d\phi^2 , \label{gamma}
\end{eqnarray}
where $I$ and $J$ denote $1$ and $2$ 
and we used Eq. (\ref{ds2-SSS-AB}). 
We refer to $\gamma_{IJ}$ as the optical metric. 
The optical metric can be used to describe 
a two-dimensional Riemannian space. 
This Riemannian space is denoted as $M^{\mbox{opt}}$. 
The light ray is a spatial geodetic curve in $M^{\mbox{opt}}$.

%\subsection{Angles}
In the optical metric space $M^{\mbox{opt}}$, 
let $\Psi$ denote the angle between the light propagation direction and the radial direction. 
A straightforward calculation gives   
\begin{eqnarray}
\cos \Psi = 
\frac{b \sqrt{A(r)B(r)}}{r^2} \frac{dr}{d\phi} . 
\label{cosPsi}
\end{eqnarray}
This is rewritten as 
\begin{equation}
\sin\Psi = \frac{b \sqrt{A(r)}}{r} , 
\label{sinPsi}
\end{equation}
where we used Eq. (\ref{orbiteq}).

We denote $\Psi_R$ and $\Psi_S$ as 
the directional angles of the light propagation.  
$\Psi_R$ and $\Psi_S$are measured 
at the receiver position (R) and the source position  (S), respectively. 
We denote $\phi_{RS} \equiv \phi_R - \phi_S$  
the coordinate separation angle 
between the receiver and source. 
By using these angles 
$\Psi_R$, $\Psi_S$ and $\phi_{RS}$, 
we define 
\begin{equation}
\alpha \equiv \Psi_R - \Psi_S + \phi_{RS} . 
\label{alpha}
\end{equation}
This is a basic tool that was invented in Reference \cite{Ishihara2016}. 
In the following, we shall prove that 
the definition by Eq. (\ref{alpha}) is geometrically  invariant \cite{Ishihara2016, Ishihara2017}.

\begin{figure}
\includegraphics[width=15cm]{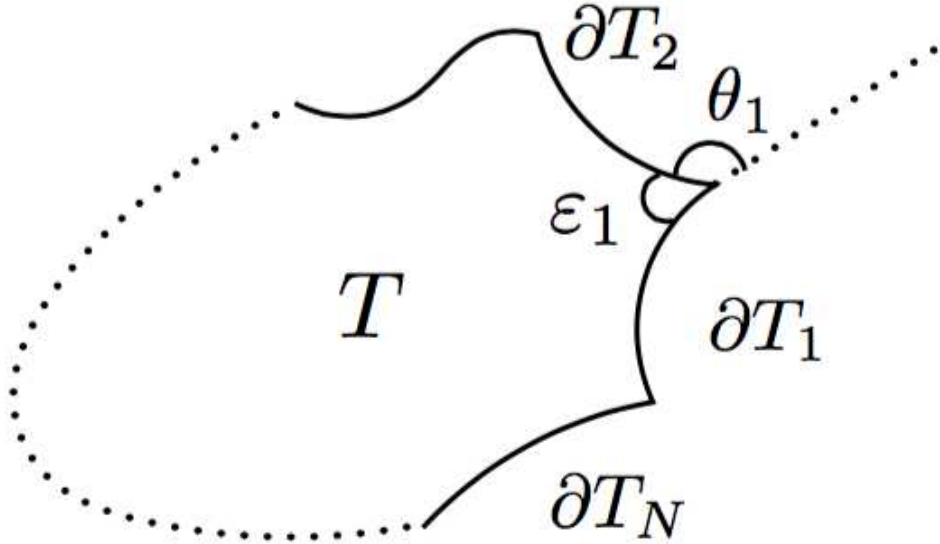}
\caption{
Gauss-Bonnet theorem. 
We consider a closed curve in a surface. 
}
\label{fig-GB}
\end{figure}

Here, we briefly mention the Gauss-Bonnet theorem.  
$T$ is a two-dimensional orientable surface.
Differentiable curves $\partial T_a$ ($a=1, 2, \cdots,  N$) are its boundaries. 
Please see Figure \ref{fig-GB} for the orientable surface. 
We denote the jump angles between the curves 
as $\theta_a$ 
($a=1, 2, \cdots, N$). 
The Gauss-Bonnet theorem: 
\cite{GB-theorem}
\begin{eqnarray}
\iint_{T} K dS + \sum_{a=1}^N \int_{\partial T_a} \kappa_g d\ell + 
\sum_{a=1}^N \theta_a = 2\pi , 
\label{localGB}
\end{eqnarray}
where 
$\ell$ means the line element of the boundary curve, 
$dS$ denotes the area element of the surface, 
$K$ means the Gaussian curvature of the surface $T$, 
$\kappa_g$ is the geodesic curvature of $\partial T_a$. 
The sign of $\ell$ is chosen to be consistent 
with the surface orientation.

\begin{figure}
\includegraphics[width=15cm]{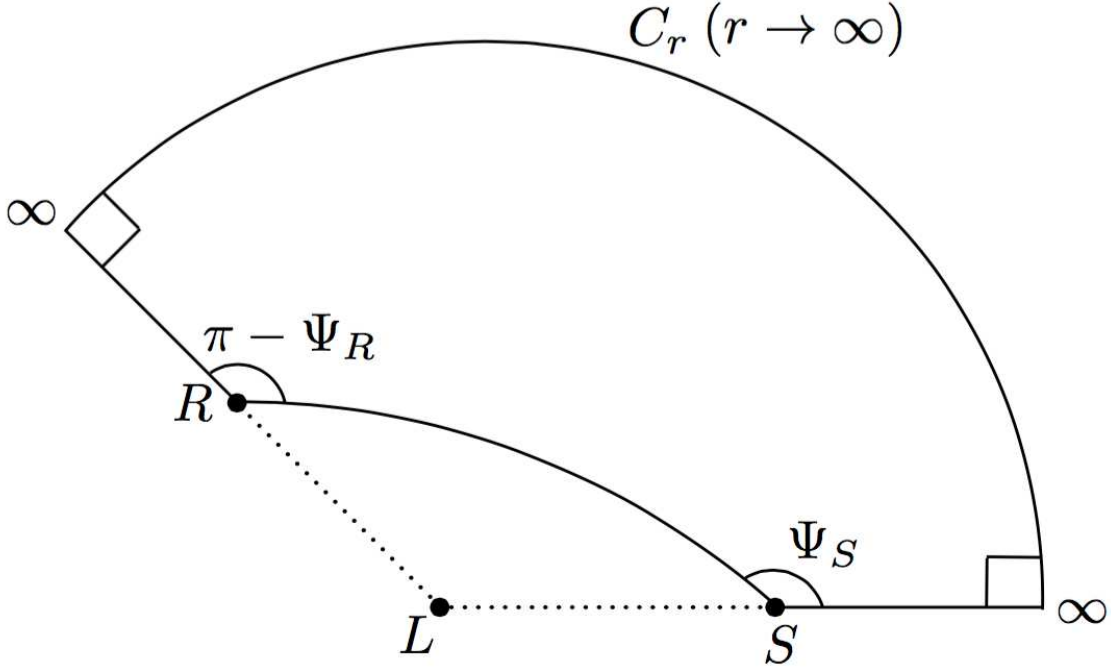}
\caption{
${}^{\infty}_{R}{\square}^{\infty}_{S}$ 
is a quadrilateral embedded in a curved space. 
}
\label{fig-Box}
\end{figure}

Suppose a quadrilateral 
${}^{\infty}_{R}{\square}^{\infty}_{S}$. 
Please see Figure \ref{fig-Box} for this. 
This is made of four lines; 
(1) the spatial curve for the light ray, 
(2) and (3) two outgoing radial lines from R and from S 
and (4) a circular arc segment $C_r$ 
that is centered at the lens 
with the coordinate radius $r_C$ ($r_C \to \infty$) 
and intersects the radial lines  
at the receiver or the source. 
We restrict ourselves within 
the asymptotically flat spacetime. 
Then, 
$\kappa_g \to 1/r_C$ and $d\ell \to r_C d\phi$ 
as $r_C \to \infty$ (See e.g. \cite{GW2008}). 
By using them, we find 
$\int_{C_r} \kappa_g d\ell \to \phi_{RS}$.  
Applying this result to the Gauss-Bonnet theorem 
for ${}^{\infty}_{R}{\square}^{\infty}_{S}$, 
we obtain 
\begin{eqnarray}
\alpha 
&=& \Psi_R - \Psi_S + \phi_{RS} 
\nonumber\\
&=& - \iint_{{}^{\infty}_{R}{\square}^{\infty}_{S}} K dS . 
\label{alpha-K}
\end{eqnarray}
Therefore, $\alpha$ is shown to be 
invariant for transformations of the spatial  coordinates.  
In addition, 
$\alpha$ is well-defined 
even when $L$ is a singular point. 
This is because the point $L$ 
does not appear in the surface integral nor 
in the line integral. 
Furthermore, $\alpha$ vanishes in Euclidean space. 
This means $\alpha$ is a measure 
of the deviation from the flat space.

\begin{figure}
\includegraphics[width=15cm]{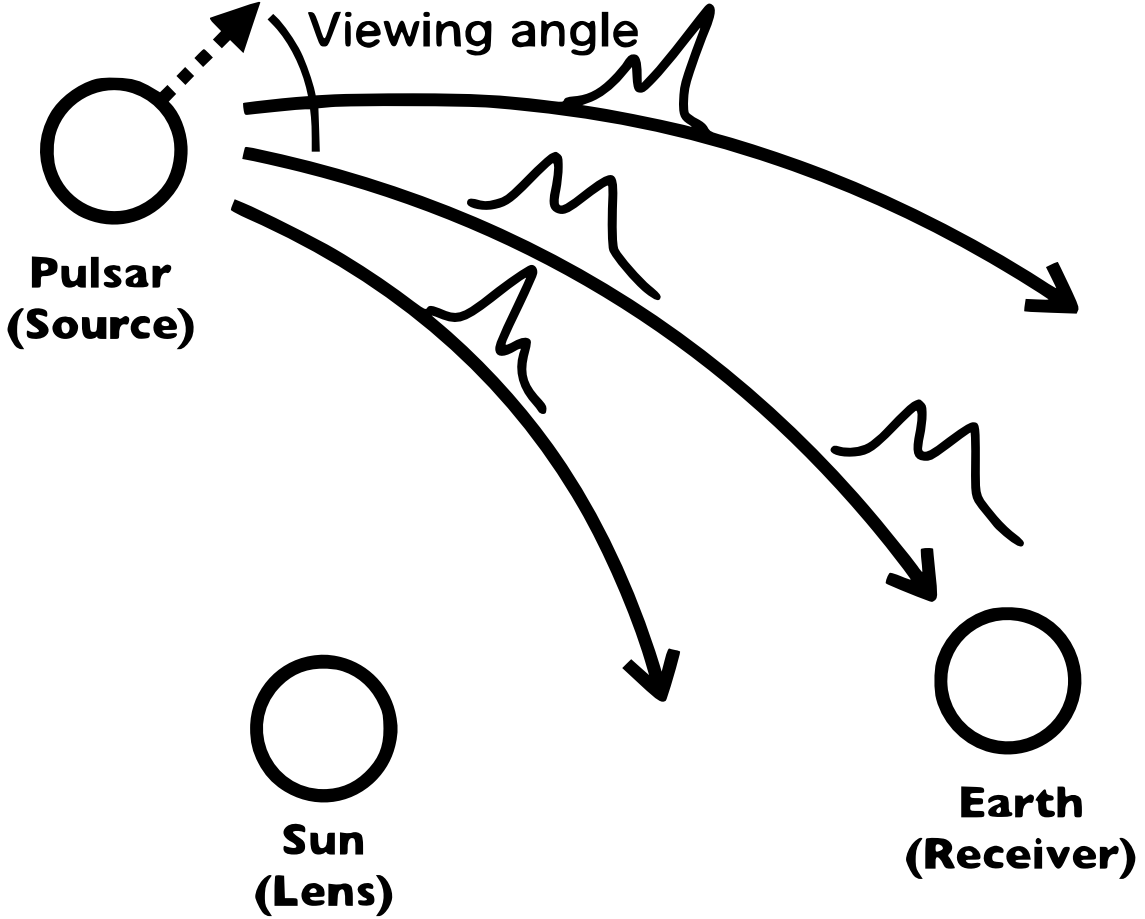}
\caption{
Observable $\alpha$ in Eq. (\ref{alpha}). 
In this schematic figure, the lens, receiver and source are 
the Sun, the Earth and a pulsar that periodically radiates radio signals in a specific anistropic manner. From the pulse profile, we can determine the radiation direction at the source. By using the ephemeris, we know the relative 
positions of the Sun, Earth and the pulsar. 
Hence, we can determine $\phi_{RS}$ and $\Psi_S$. 
By observing the pulsar, we can measure $\Psi_R$. 
In principle, therefore, 
we can determine $\Psi_R - \Psi_S + \phi_{RS}$ 
from these astronomical observations. 
}
\label{fig-Alpha}
\end{figure}

Here, we explain that $\alpha$ defined in Eq. (\ref{alpha}) is observable in principle. 
For the simplicity, let us imagine the following ideal situation. 
The positions of a source and receiver are known. 
For instance, we assume that the lens object is the Sun, the receiver is located at the Earth, 
and the source is a pulsar which radiates radio signals with a constant period in an anisotropic manner. In particular, we assume that the source is one of the known pulsars whose spin period 
and pulse signal behaviors such as pulse profiles are well-understood. 
By very accurate radio observations such as VLBI, the relative positions of 
the Earth, Sun and the pulsar can be determined from the ephemeris. 
(1) From this, we can know $\phi_{RS}$ in principle. 
(2) We can directly measure  the angle $\Psi_R$ at the Earth 
between the solar direction and the pulsar direction.   
(3) More importantly, the direction of radiating the pulses that reaches the receiver 
can be also determined in principle, because the viewing angle of the pulsar seen by the receiver is known from the pulse profiles. The viewing angle is changing with time  because of the Earth motion around the Sun.  By using the pulsar position and the pulse radiation direction, we can determine $\Psi_S$. 
Please see Figure 3 for this situation. 
We explain in more detail how $\Psi_S$ at S can be measured by the observer at R. We consider a pulsar whose spin axis is known from some astronomical observations. 
A point is that the spin axis of an isolated pulsar is constant with time. 
The pulse shape and profile depend on the viewing angle with respect to the spin axis of the pulsar. 
The Earth moves around the Sun and hence the observer sees the same pulsar with different viewing angles with time. 
Accordingly, the observed pulse shape changes. 
By observing such a change in the pulse shape, we can in principle 
determine the intrinsic direction of the radio emission, namely the angle between the spin axis and the direction of the emitted light to the observer. 
In addition, we can know the intrinsic position (including the radial direction from the lens) of such a known pulsar from the ephemeris. By using the intrinsic position (its radial direction) and emission direction at S, $\Psi_S$ can be determined in principle, though it is very difficult with current technology. 
As a result, we can determine in principle $\Psi_R - \Psi_S + \phi_{RS}$ from astronomical observations. Namely, $\alpha$ in Eq. (\ref{alpha}) is observable. 
Note that this procedure does not need assume a different spacetime, 
while such a fiducial spacetime was assumed by Arakida (2018) \cite{Arakida2018}, though 
the receiver in our universe cannot observe the fiducial different spacetime 
but can assume (or make theoretical calculations of) some quantities on the different spacetime.

One can easily see that, 
in the far limit of 
the source and the receiver,  
Eq. (\ref{alpha-K}) agrees with 
the deflection angle of light as 
\begin{equation}
\alpha_{\infty} = 2 \int_0^{u_0} \frac{du}{\sqrt{F(u)}} - \pi .  
\label{alpha-infty}
\end{equation}
Here, we define 
$u$ and $u_0$ as 
as the inverse of $r$, 
the inverse of the closest approach 
(often denoted as $r_0$), respectively.  
$F(u)$ is defined as 
\begin{eqnarray}
F(u) \equiv \left( \frac{du}{d\phi} \right)^2 . 
\label{F}
\end{eqnarray}
$F(u)$ can be computed by using Eq. (\ref{orbiteq}). 

The present paper wishes to avoid 
the far limit in the following reason. 
Every observed stars and galaxies are never 
located at infinite distance from us. 
For instance, we observe finite-redshift galaxies 
in cosmology. 
We cannot see objects at infinite redshift 
(exactly at the horizon). 
Except for a few rare cases in astronomy, 
the distance to the light source is much larger 
than the size of the lens. 
Therefore, we find a strong motivation for studying 
a situation that 
the distance from the source to the receiver is finite. 
We define $u_R$ and $u_S$ as the inverse of 
$r_R$ and $r_S$, respectively, 
where $r_R$ and $r_S$ are finite. 
Eq. (\ref{alpha}) is rewritten in an explicit form as 
\cite{Ishihara2016, Ishihara2017} 
\begin{equation}
\alpha = \int_{u_R}^{u_0} \frac{du}{\sqrt{F(u)}} 
+ \int_{u_S}^{u_0} \frac{du}{\sqrt{F(u)}} 
+\Psi_R - \Psi_S . 
\label{alpha-finite}
\end{equation}
Here, we assume light rays that have the only one local minimum of the radius coordinate between $r_S$ and $r_R$. This is valid for normal situations in astronomy. 
However, we should note that multiple local minima are possible, e.g. if the emitter or the receiver (or both) are between the horizon and the light sphere in the Schwarzschild spacetime, or if the emitter and receiver are at different sides of the throat of a wormhole spacetime. For such a case of multiple local minima, Eq. (12) has to be modified, because it assumes only the local minimum at $u = u_0$.

\section{Weak deflection of light in Schwarzschild spacetime}
In this section, we consider the weak deflection of light in Schwarzschild spacetime, 
for which the line element becomes 
\begin{eqnarray}
ds^2 &=& -\left( 1 - \frac{r_g}{r} \right) 
dt^2 
+ \frac{dr^2}{\displaystyle 1 - \frac{r_g}{r}} 
\nonumber\\
&&
+ r^2 (d\theta^2 + \sin^2\theta d\phi^2) ,  
\label{ds-Sch}
\end{eqnarray}
where $r_g = 2M$ in the geometrical unit. 
Then, $F(u)$ is 
\begin{equation}
F(u) = \frac{1}{b^2} - u^2 + r_g u^3 . 
\label{F-Sch}
\end{equation}

By using Eq. (\ref{sinPsi}), 
$\Psi_R - \Psi_S$ in the Schwarzschild spacetime is expanded as 
\begin{eqnarray}
\Psi_R^{\mbox{Sch}} - \Psi_S^{\mbox{Sch}} 
&\equiv& [\arcsin(bu_R) + \arcsin(bu_S) - \pi] 
\nonumber\\
&& - \frac12 b r_g 
\left( \frac{u_R^2}{\sqrt{1 - b^2u_R^2}}
+ \frac{u_S^2}{\sqrt{1 - b^2u_S^2}} \right) 
+ O(b r_g^2 u_S^3, b r_g^2 u_R^3) . 
\label{Psi-Sch}
\end{eqnarray}
Note that $\Psi_R - \Psi_S \to \pi$ 
in the Schwarzschild  spacetime  
as $u_S \to 0$ and $u_R \to 0$.

\section{Other examples}
This section discusses two examples for 
a non-asymptotically flat spacetime. 
One is the Kottler solution to the Einstein equation. 
The other is an exact solution in the Weyl conformal  gravity. 
The aim of this study is 
to give us a suggestion or a speculation. 
We note that the present formulation is 
limited within the asymptotic flatness, 
rigorously speaking. 
As mentioned in Introduction, 
Arakida \cite{Arakida2018} made an attempt to apply the Gauss-Bonnet theorem to quadrilaterals that are not extending to infinity, though a comparison between two different manifolds that he proposed is an open issue. 
A more careful study that gives a justification 
for this speculation or perhaps disprove it 
will be left for future. 

In this section, we do not assume 
the source at the past null infinity ($r_S \to \infty$)
nor the receiver at the future null infinity ($r_R \to \infty$), 
because $A(r)$ diverges or does not exist as $r \to \infty$. 
We keep in mind that the source and receiver 
are located at finite distance from the lens object. 
Therefore, we use Eq. (\ref{alpha-finite}). 
As mentioned already,  
Eq. (\ref{sinPsi}) is more useful for calculating  $\Psi_R$ and $\Psi_S$ 
than Eq. (\ref{cosPsi}), 
because Eq. (\ref{sinPsi}) requires only the local quantities but not any differentiation.  
By straightforward calculations, we obtain 
the following results for the above two models. 

\subsection{Kottler solution}
We consider the Kottler solution \cite{Kottler}. 
This solution is written as 
\begin{eqnarray}
ds^2 &=& -\left( 1 - \frac{2M}{r} - \frac{\Lambda}{3}r^2 \right) 
dt^2 
+ \frac{dr^2}{\displaystyle 1 - \frac{2M}{r} - \frac{\Lambda}{3}r^2} 
\nonumber\\
&&
+ r^2 (d\theta^2 + \sin^2\theta d\phi^2) , 
\label{ds-Kottler}
\end{eqnarray}
where the cosmological constant is denoted 
by $\Lambda$. 

We use Eq. (\ref{sinPsi}), such that  
$\Psi_R - \Psi_S$ can be expanded in terms of $2M$ and $\Lambda$ as 
\begin{align}
\Psi_R-\Psi_S
&=\Psi^{Sch}_R-\Psi^{Sch}_{S} 
- \frac{b\Lambda}{6u_R\sqrt{1-b^2u_R^2}} 
- \frac{b\Lambda}{6u_S\sqrt{1-b^2u_S^2}} 
\notag\\
&+ \frac{bu_R(-1+2b^2u_R^2)}{8(1-b^2u_R^2)^{3/2}}
\left(4M^2u_R^2+\frac{4M\Lambda}{3u_R}+\frac{\Lambda^2}{9u_R^4}\right)
\notag\\
&+\frac{bu_S(-1+2b^2u_S^2)}{8(1-b^2u_S^2)^{3/2}}
\left(4M^2u_S^2+\frac{4M\Lambda}{3u_S}
+\frac{\Lambda^2}{9u_S^4}\right) 
\notag\\
&+ O(M^3, M^2\Lambda, M\Lambda^2, \Lambda^3) . 
\label{Psi-Kottler}
\end{align}
Here, $\Psi_R^{\mbox{Sch}} - \Psi_S^{\mbox{Sch}}$ 
is a pair of the terms that appear also 
in a case of the Schwarzschild spacetime.  
The above expansion of $\Psi_R - \Psi_S$ has 
a divergent term 
in the limit as $u_S \to 0$ and $u_R \to 0$. 
The reason for this divergent behavior is 
that the spacetime is not asymptotically flat and therefore 
the limit of $u_S \to 0$ and $u_R \to 0$ is no longer  allowed. 
Hence, the power series in Eq. (\ref{Psi-Kottler}) 
is mathematically valid 
only within a convergence radius. 

For the Kottler spacetime, $F(u)$ becomes 
\begin{equation}
F(u) = \frac{1}{b^2} - u^2 + r_g u^3 
+ \frac{\Lambda}{3} . 
\label{F-Kottler}
\end{equation}
We obtain 
\begin{align}
\phi_{RS}
=&\pi-\arcsin(bu_{R})-\arcsin(bu_{S})
\notag\\
&+\frac{r_{g}}{b}\left[\frac{1}{\sqrt{1-b^{2}u_{R}^{2}}} 
\left(1-\frac{1}{2}b^{2}u_{R}^{2}\right) 
+ \frac{1}{\sqrt{1-b^{2}u_{S}^{2}}}\left(1-\frac{1}{2}b^{2}u_{S}^{2}\right)\right]
\notag\\
&+\frac{\Lambda b^{3}}{6}\left[\frac{u_{R}}{\sqrt{1-b^{2}u_{R}^{2}}} 
+ \frac{u_{S}}{\sqrt{1-b^{2}u_{S}^{2}}}\right] 
+ \frac{r_{g}\Lambda b}{12}
\left[\frac{2-3b^{2}u_{R}^{2}}{(1-b^{2}u_{R}^{2})^{\frac{3}{2}}}
+\frac{2-3b^{2}u_{S}^{2}}{(1-b^{2}u_{S}^{2})^{\frac{3}{2}}}\right]
+ O(r_{g}^{2},\Lambda^{2}) .
\label{phi-Kottler}
\end{align}

By using Eqs. (\ref{Psi-Kottler}) and (\ref{phi-Kottler}), 
$\alpha$ is obtained as 
\begin{align}
\alpha
=&
\frac{r_{g}}{b}\left[\sqrt{1-b^{2}u_{R}^{2}}+\sqrt{1-b^{2}u_{S}^{2}}\right]
\notag\\
& -\frac{\Lambda b}{6}
\left[\frac{\sqrt{1-b^{2}u_{R}^{2}}}{u_{R}}+\frac{\sqrt{1-b^{2}u_{S}^{2}}}{u_{S}}\right]
\notag\\
& +\frac{r_{g}\Lambda b}{12}\left[\frac{1}{\sqrt{1-b^{2}u_{R}^{2}}}
+\frac{1}{\sqrt{1-b^{2}u_{S}^{2}}}\right] 
+ O(r_{g}^{2},\Lambda^{2}) . 
\label{alpha-Kottler}
\end{align}
This equation has 
several divergent terms 
as $b u_R \to 0$ and $b u_S \to 0$. 
The apparent divergent is problematic only in 
the case that the source or receiver is located 
at the horizon. 
In other words, all the terms 
in Eq. (\ref{alpha-Kottler}) are finite 
and thus harmless 
for astronomical situations.

\subsection{Weyl conformal gravity case}
Next, we consider Weyl conformal gravity model. 
This theory was originally suggested by Bach 
\cite{Bach}.
The SSS solution in this model is expressed by  introducing three new parameters that 
are often denoted as $\beta$, $\gamma$ and $k$. 
For this generalized solution in conformal gravity, 
Birkhoff's theorem still holds 
\cite{Riegert}. 
The SSS solution in the Weyl gravity model is 
\cite{MK}
\begin{eqnarray}
ds^2&=&-A(r) dt^2+\frac{1}{A(r)} dr^2 
+ r^2(d\theta^2+\sin^2\theta d\phi^2), \nonumber\\
A(r)&=&1 - 3m\gamma - \frac{2m}{r} + \gamma r -k r^2 .
\label{ds-Weyl}
\end{eqnarray}
Here, 
$m \equiv \beta (2 - 3 \beta\gamma)/2$. 
$kr^2$ in the metric plays the same role 
as the cosmological constant in the Kottler spacetime 
that has been studied above. 
Therefore, 
we omit the $r^2$ term for simplifying our analysis. 

By using Eq. (\ref{sinPsi}), we expand 
$\Psi_R - \Psi_S$ in $\beta$ and $\gamma$. 
The result is  
\begin{align}
\Psi_R-\Psi_S \equiv &\Psi_R^{\mbox{Sch}}-\Psi_S^{\mbox{Sch}} \notag\\
&+\frac{b\gamma}{2}\left(\frac{u_R}{\sqrt{1-b^2u^2_R}}+\frac{u_S}{\sqrt{1-b^2u^2_S}}\right) \notag\\
&-\frac{m\gamma}{2}\left[\frac{bu_R(2-b^2u_R^2)}{(1-b^2u_R^2)^{3/2}}+\frac{bu_S(2-b^2u_S^2)}{(1-b^2u_S^2)^{3/2}}\right] +O(m^2,\gamma^2) . 
\label{Psi-Weyl}
\end{align}
We should note that this expansion of 
$\Psi_R - \Psi_S$ is divergent 
as $u_S \to 0$ and $u_R \to 0$.  
This divergent behavior is not so problematic, 
because the limit of $u_S \to 0$ and $u_R \to 0$ 
is not allowed in this spacetime. 
Hence, we note that, rigorously speaking, 
Eq. (\ref{Psi-Weyl}) 
is mathematically valid 
only within a convergence radius. 

For the present case omitting $k$, 
we obtain 
\begin{equation}
F(u) = \frac{1}{b^2} - u^2 + 2m u^3 
+ \gamma u^2 - \gamma u . 
\label{F-Weyl}
\end{equation}
$\phi_{RS}$ is computed as 
\begin{align}
\phi_{RS}
=& [\pi-\arcsin(bu_R)-\arcsin(bu_S)] 
\notag\\
&
+ \frac{m}{b} \left( \frac{2-b^2u_R^2}{\sqrt{1-b^2u_R^2}} 
+ \frac{2-b^2u_S^2}{\sqrt{1-b^2u_S^2}} \right)
\notag\\
& - \frac{\gamma}2 \left( \frac{b}{\sqrt{1-b^2u_R^2}} + \frac{b}{\sqrt{1-b^2u_R^2}} \right) 
\notag\\ 
& 
+\frac{m\gamma}{2}\left[\frac{b^3u_R^3}{(1-b^2u_R^2)^{3/2}}+\frac{b^3u_S^3}{(1-b^2u_S^2)^{3/2}}\right] 
+ O(m^2,\gamma^2) .
\label{phi-Weyl}
\end{align}

Consequently, we obtain $\alpha$ as 
\begin{align}
\alpha=
& \frac{2m}{b} \left(\sqrt{1-b^2u_R^2}+\sqrt{1-b^2u_S^2}\right) 
\notag\\
& -m\gamma\left(\frac{bu_R}{\sqrt{1-b^2u_R^2}}+\frac{bu_S}{\sqrt{1-b^2u_S^2}}\right) + O(m^2,\gamma^2) . 
\label{alpha-Weyl}
\end{align}
The linear terms in $\gamma$ cancel out 
with each other and they do not appear in the final expression for the deflection angle of light. 
This result may suggest a correction 
to the results in previous papers 
\cite{Edery,Sultana,Cattani} 
that reported non-zero contributions from $\gamma$.

\subsection{Far source and receiver}
Next, we investigate a situation 
of a distant source and receiver from 
the lens object: 
$b u_S \ll 1$ and $b u_R \ll 1$. 
Divergent terms in the deflection angle appear 
in the limit as $b u_S \to 0$. 
Therefore, We carefully investigate 
the leading part 
in a series expansion, where 
the infinite limit is not taken. 
As a result, approximate expressions 
for the deflection of light 
are obtained as follows. 

\noindent
(1) Kottler model:\\
The expression for $\phi_{RS}$ in this 
approximation is the same as 
the seventh and eighth terms of Eq. (5) in \cite{Sereno}, 
the third and fifth terms of Eq. (15) in \cite{Bhadra}, 
and the second term of Eq. (14) in \cite{ArakidaK}. 
On the other hand, 
they \cite{Sereno,Bhadra,ArakidaK} did not 
take account of $\Psi_R - \Psi_S$. 
In the far approximation, 
Eq. (\ref{alpha-Kottler}) becomes 
\begin{equation}
\alpha \sim \frac{4M}{b} 
- \frac16 \Lambda b \left( \frac{1}{u_R} + \frac{1}{u_S} \right) 
+ \frac{1}{3} M \Lambda b . 
\label{alpha-far-Kottler}
\end{equation}
This expression suggestions a correction 
to the earlier works \cite{Sereno,Bhadra,ArakidaK}.  
For instance, only the term of $\phi_{RS}$ 
was considered in Sereno (2009). 
%Please see \cite{previous} for a more subtle case associated 
%with Rindler and Ishak's approach. 

\noindent
(2) Weyl conformal gravity model:\\
Next, we consider 
the Weyl conformal gravity model. 
The deflection angle of light 
in the far approximation is computed as 
\begin{eqnarray}
\alpha
&\sim& 
\frac{4m}{b}  
+ O(m^2,\gamma^2) , 
\label{alpha-Weyl-far}
\end{eqnarray}
where $m\gamma$ parts from $\Psi_R - \Psi_S$ and from $\psi_{RS}$ 
cancel out with each other. 
Please see also 
Eqs. (\ref{Psi-Weyl}) and (\ref{phi-Weyl}).  
For instance, Reference \cite{Lim2017} gives the exact expression of the deflection angle for the asymptotic receiver and  source in the Kottler and Weyl conformal gravity spacetime.

\section{Extension to the Strong deflection of light}
In the previous sections, we considered the weak deflection of light: 
A light ray from the source to the receiver 
is expressed by a spatial curve. 
The curve is simply-connected. 
In the strong deflection limit, on the other hand, 
it is possible that the spatial curve has a winding  number with intersection  points. 
We thus divide the whole curve into segments. 
And it is easier to investigate each simple segment. 

\subsection{Loops in the photon orbit}
\begin{figure}
\includegraphics[width=14cm]{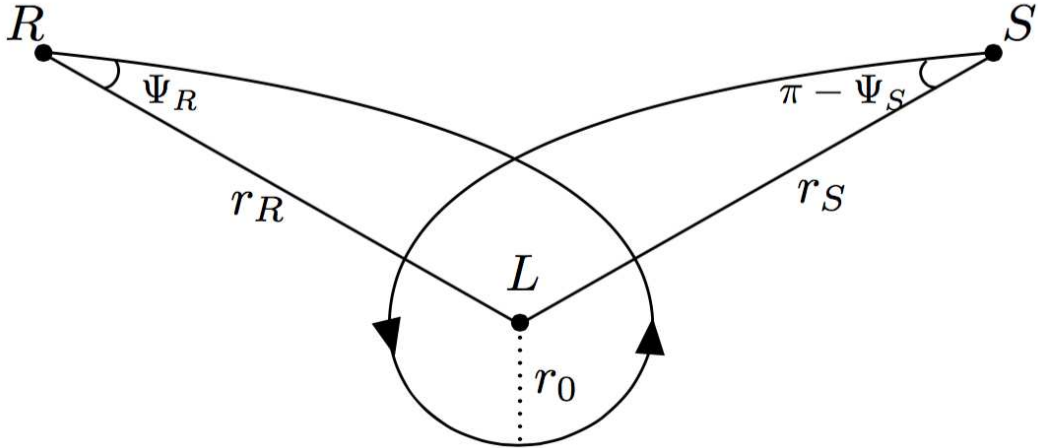}
\caption{
A one-loop case 
for the photon orbit 
in $M^{\mbox{opt}}$. 
}
\label{fig-oneloop}
\end{figure}

We begin with one loop case of the light ray curve. 
This case is shown by Figure \ref{fig-oneloop}. 

\begin{figure}
\includegraphics[width=10cm]{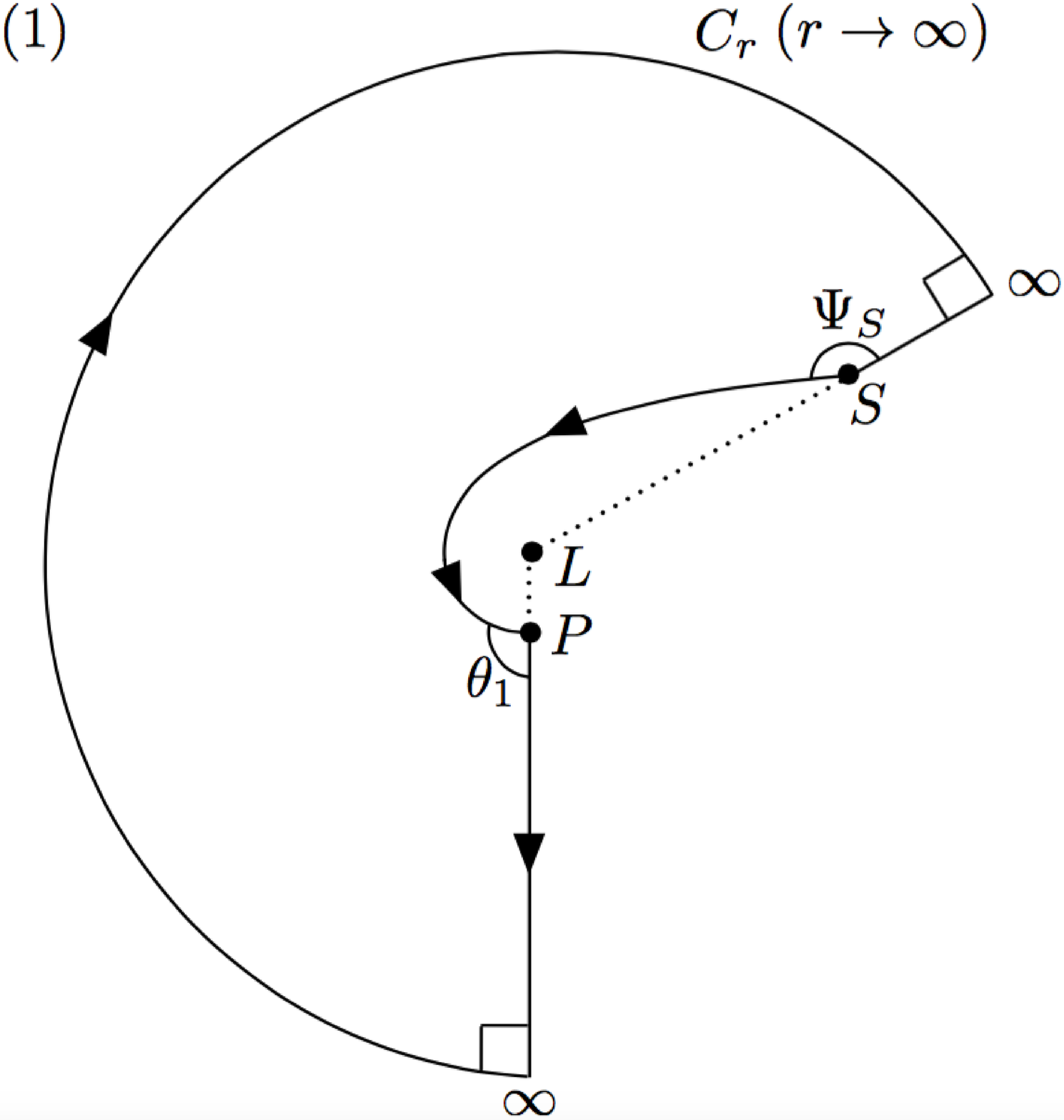}\\
\includegraphics[width=10cm]{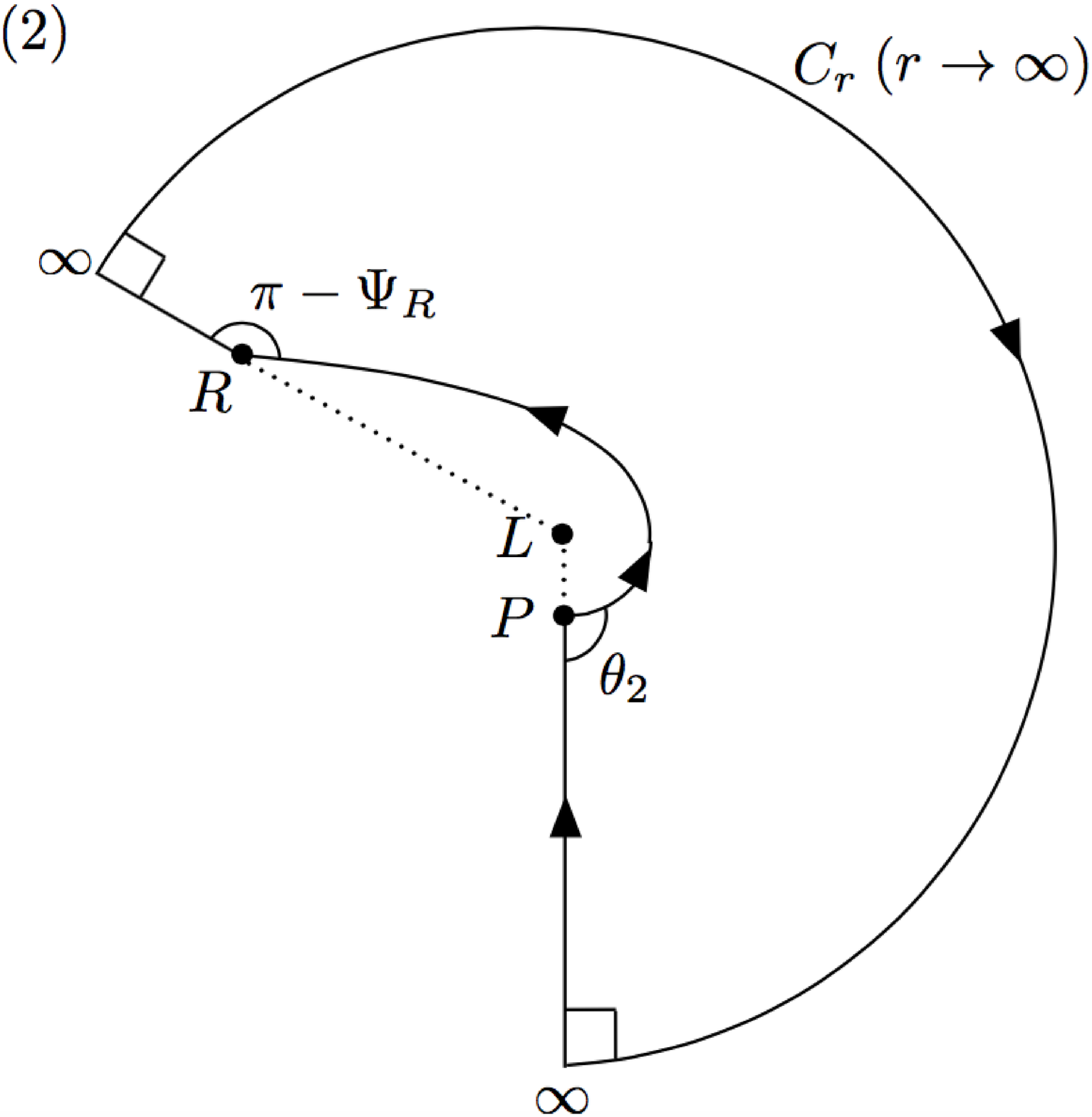}
\caption{ 
Quadrilaterals. They are made from the photon orbit 
in a non-Euclidean space. 
See Figure \ref{fig-oneloop}. 
}
\label{fig-oneloop2}
\end{figure}

First, we consider the two quadrilaterals (1) and (2) 
in Figure \ref{fig-oneloop2}. 
They can be constructed by introducing an auxiliary point (P) 
and 
next by adding auxiliary outgoing radial lines 
(solid line in this figure) from the point P 
in the quadrilaterals (1) and (2). 
The point P does not need to be the periastron. 
The direction of the two auxiliary lines in (1) and (2) 
is opposite to each other. 
The two auxiliary lines thus cancel out to 
make no contributions to $\alpha$. 
Here, $\theta_1$ and $\theta_2$ denote the inner angle 
at the point P in the quadrilateral (1) 
and that in the quadrilateral (2), respectively. 
We can see that $\theta_1 + \theta_2 = \pi$. 
This is because the line from the source 
to the receiver is a geodesic and 
the point P is located in this line. 

\begin{figure}
\includegraphics[width=14cm]{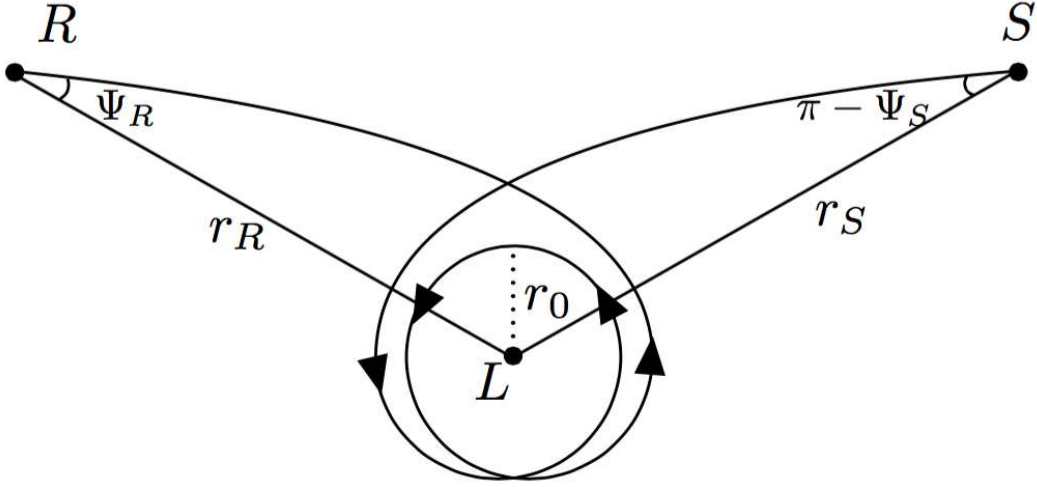}
\caption{ 
Two loops for the light ray 
in $M^{\mbox{opt}}$. 
}
\label{fig-twoloop}
\end{figure}

\begin{figure}
\includegraphics[width=7cm]{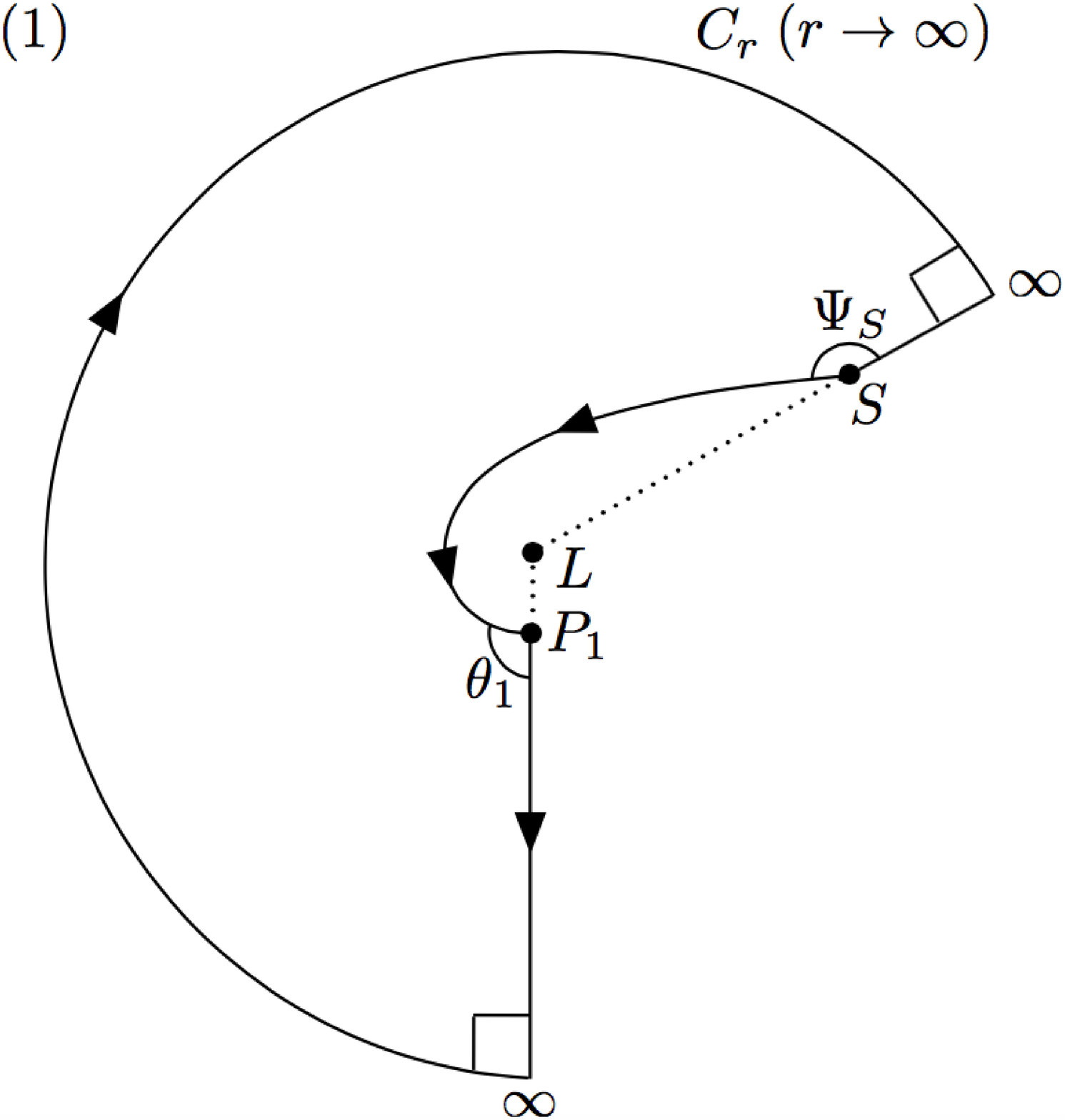}
\includegraphics[width=8cm]{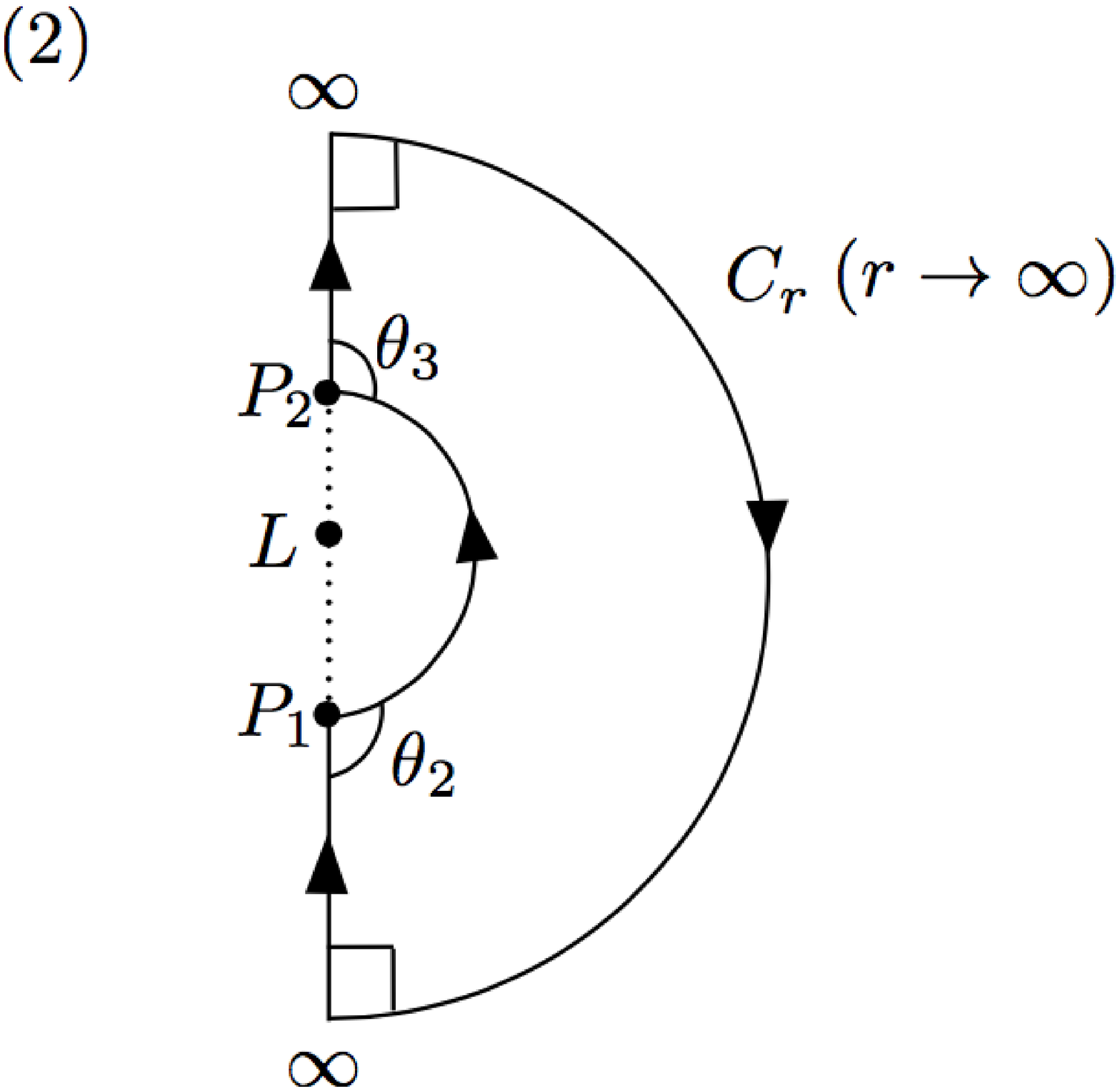}
\includegraphics[width=6cm]{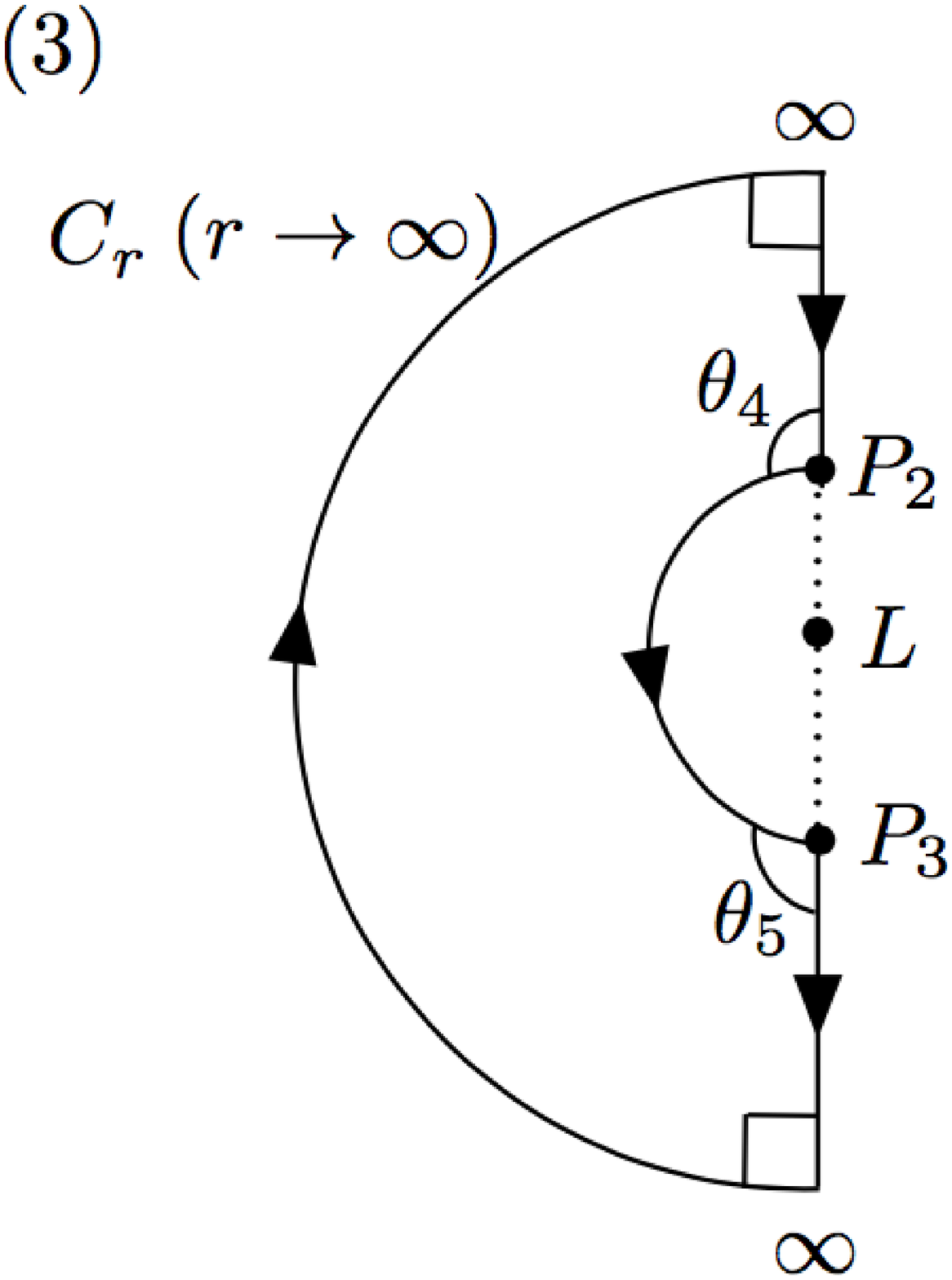}
\includegraphics[width=7cm]{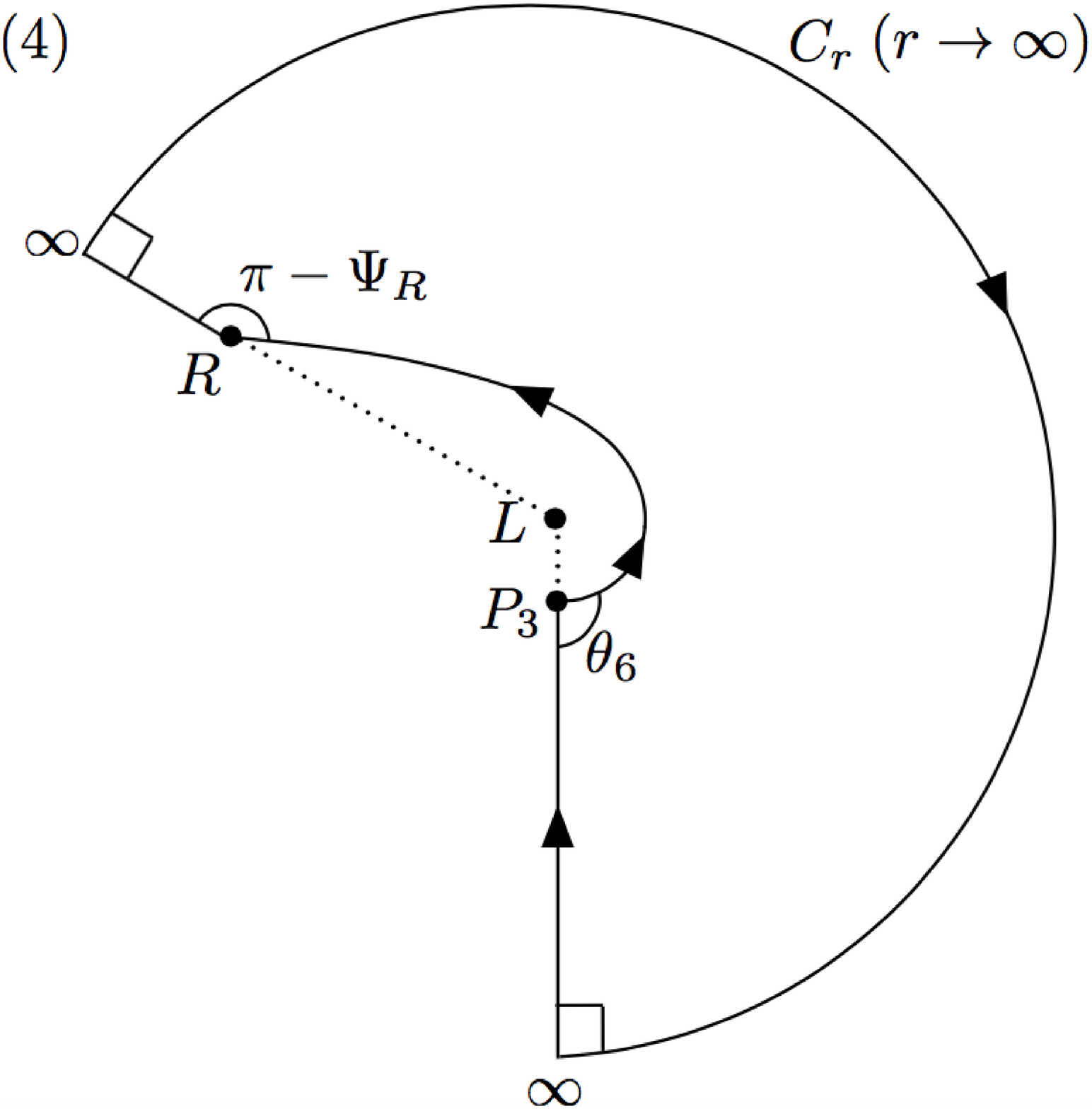}
\caption{ 
Quadrilaterals (1)-(4). 
They are in a non-Euclidean plane $M^{\mbox{opt}}$. 
See also Figure \ref{fig-twoloop}. 
}
\label{fig-twoloop2}
\end{figure}

\begin{figure}
\includegraphics[width=15cm]{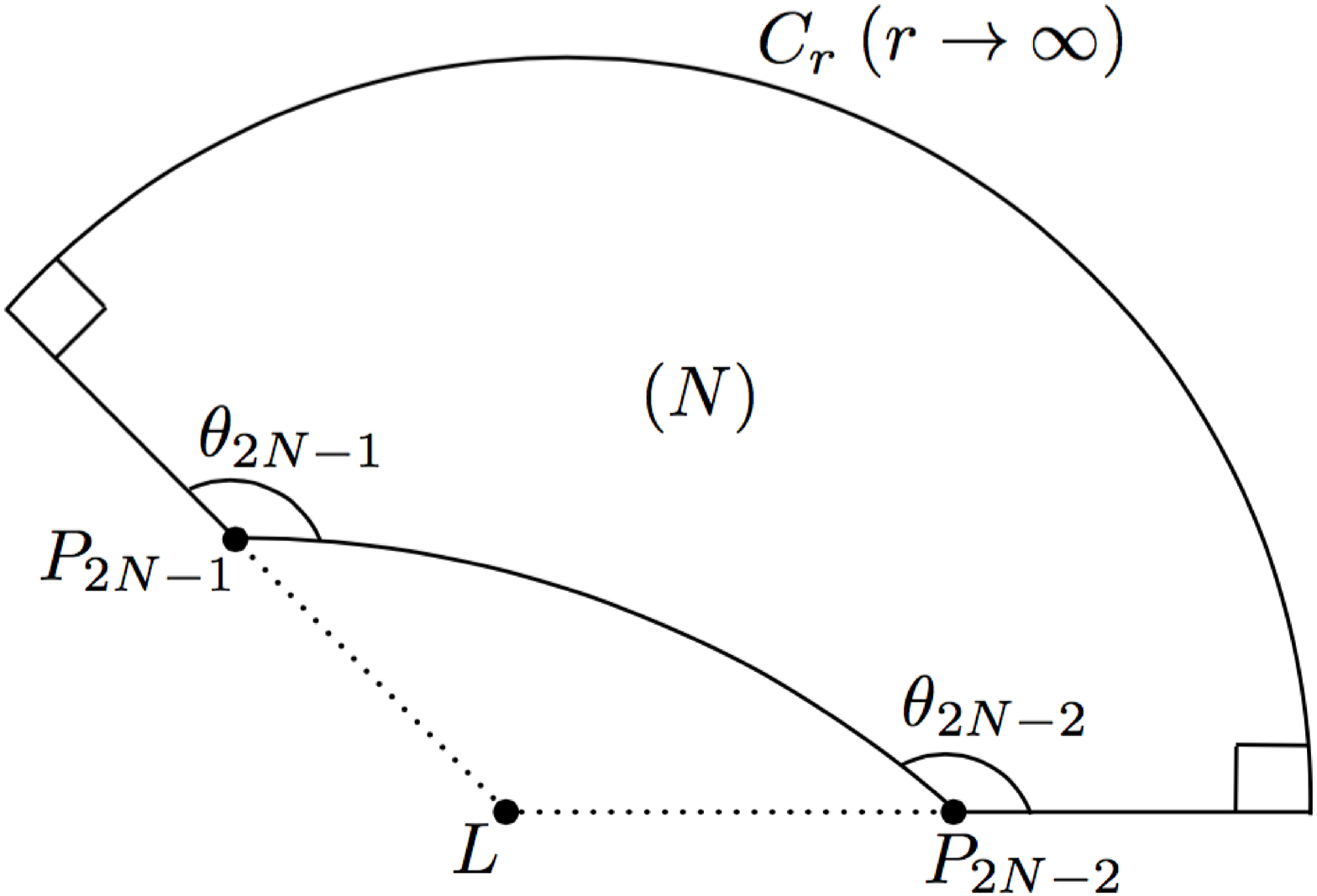}
\caption{ 
A quadrilateral in any loop number. 
This case is discussed 
when we prove by induction that Eq. (\ref{alpha}) 
holds in any loop number. 
}
\label{fig-anyloop}
\end{figure}

For a quadrilateral in Figure \ref{fig-oneloop2}, 
the method in Section II is still applicable. 
By the same way of obtaining Eq. (\ref{alpha-K}), 
we obtain 
\begin{align}
\alpha^{(1)} &= 
(\pi - \theta_1) - \Psi_S + \phi_{RS}^{(1)} , 
\nonumber\\
\alpha^{(2)} &= 
\Psi_R - \theta_2 + \phi_{RS}^{(2)} .   
\end{align}
Here, $\phi_{RS}$ 
is divided into two parts: 
One is 
$\phi_{RS}^{(1)}$ for one quadrilateral 
and the other is 
$\phi_{RS}^{(2)}$ for the other quadrilateral. 

If $r_S = r_R$, the quadrilaterals (1) and (2) 
are symmetric for reflection 
and 
$\phi_{RS}^{(1)} = \phi_{RS}^{(2)} = \phi_{RS}/2$. 
If not, $\phi_{RS}^{(1)}$ is not the same as  $\phi_{RS}^{(2)}$. 
In any case, however, $\phi_{RS}^{(1)} + \phi_{RS}^{(2)} = \phi_{RS}$. 
$\Psi_S$ and $(\pi - \Psi_R)$ are 
the inner angles at $S$ and $R$, respectively. 
Therefore,  
\begin{align}
\alpha &= \alpha^{(1)} + \alpha^{(2)} 
\nonumber\\
&= \Psi_R - \Psi_S + \phi_{RS} , 
\end{align}
where we use 
$\theta_1 + \theta_2 = \pi$ 
and
$\phi_{RS}^{(1)} + \phi_{RS}^{(2)} = \phi_{RS}$. 
This result is the same as Eq. (\ref{alpha}), 
though the validity domain is different.

Next, we investigate a case of two loops 
shown by Figure \ref{fig-twoloop}.  
For this case, we add lines in order to 
divide the shape into four quadrilaterals 
as shown by Figure \ref{fig-twoloop2}. 
We immediately find  
\begin{align}
\alpha^{(1)} &= 
(\pi - \theta_1) - \Psi_S + \phi_{RS}^{(1)} , 
\nonumber\\
\alpha^{(2)} &= 
(\pi- \theta_3) - \theta_2 + \phi_{RS}^{(2)} ,
\nonumber\\
\alpha^{(3)} &= 
(\pi- \theta_5) - \theta_4 + \phi_{RS}^{(3)} ,
\nonumber\\
\alpha^{(4)} &= 
\Psi_R - \theta_6 + \phi_{RS}^{{(4)}} ,   
\end{align}
where 
$\phi_{RS}^{(1)} + \phi_{RS}^{(2)} + \phi_{RS}^{(3)} + \phi_{RS}^{(4)} 
= \phi_{RS}$. 
Hence, we obtain 
\begin{align}
\alpha &= \alpha^{(1)} + \alpha^{(2)} + \alpha^{(3)} + \alpha^{(4)}
\nonumber\\
&= \Psi_R - \Psi_S + \phi_{RS} , 
\label{alpha-2loop}
\end{align}
where we use 
$\theta_1 + \theta_2 = \theta_3 + \theta_4 = \theta_5 + \theta_6 = \pi$.  
Eq. (\ref{alpha-2loop}) is obtained for the two-loop case in the same form as Eq. (\ref{alpha}). 
A loop does make the contribution to $\alpha$ 
only through the terms of $\phi_{RS}^{(2)} + \phi_{RS}^{(3)}$.

Finally, we shall complete the proof. 
We consider 
the arbitrary winding number, say $W$. 
For this case, we prepare $2W$ quadrilaterals. 
We denote 
the inner angles at finite distance from $L$ 
as $\theta_0, \cdots, \theta_{2W}$ 
in order from $S$ to $R$ 
as shown by Figure \ref{fig-anyloop}. 
Here, $\theta_0 = \Psi_S$ and $\theta_{2W} = \pi - \Psi_R$. 
Neighboring quadrilaterals (N) and (N+1) 
make the contribution to $\alpha$ only through 
$\phi_{RS}^{(N)} + \phi_{RS}^{(N+1)}$. 
We can understand this by noting that 
$\theta_{2N-1} + \theta_{2N} = \theta_{2N+1} + \theta_{2N+2} = \pi$ 
and the auxiliary lines cancel out.  
By induction, therefore, we complete the proof; 
Eq. (\ref{alpha}) holds for any winding number.

Eq. (\ref{alpha}) is equivalent to Eq. (\ref{alpha-finite}). 
This is shown by using the orbit equation. 
This expression is rearranged as 
\begin{align}
\alpha&=\Psi_{R}-\Psi_{S}+\phi_{RS}
\notag\\
&=\Psi_{R}-\Psi_{S}
+\int^0_{u_R}\frac{du}{\sqrt{F(u)}}
+\int^0_{u_S}\frac{du}{\sqrt{F(u)}}
+2\int^{u_0}_0\frac{du}{\sqrt{F(u)}} . 
\label{alpha-full}
\end{align}

We define the difference between the asymptotic deflection angle 
and the deflection angle for the finite distance case 
as $\delta\alpha$. 
\begin{align}
\delta\alpha \equiv \alpha - \alpha_{\infty}  . 
\label{delta-alpha}
\end{align}
The meaning of this is 
the finite-distance correction to the deflection angle of light. 
By substituting Eqs. (\ref{alpha-infty}) and (\ref{alpha-full}) 
into Eq. (\ref{delta-alpha}), we get 
\begin{align}
\delta\alpha = 
(\Psi_{R}-\Psi_{S}+\pi) + \int^0_{u_R}\frac{du}{\sqrt{F(u)}}
+\int^0_{u_S}\frac{du}{\sqrt{F(u)}} . 
\label{delta-alpha2}
\end{align}
This expression implies two origins of the  finite-distance corrections. 
One origin is 
$\Psi_{R}$ and $\Psi_{S}$. 
They are angles that are defined in a curved space. 
%\cite{comment-Bozza+}. 
The other origin is the two path integrals. 
They contain the information on 
the curved space. 
If we consider a receiver and source 
in the weak gravitational field 
(as common in astronomy), 
the finite-distance correction reflects 
only the weak field region, 
even if the light ray passes 
through a strong field region.

\section{Strong deflection of light in Schwarzschild spacetime} 
In this section, we consider the Schwarzschild black hole. 
%Then, Eq. (\ref{F}) becomes 
%\begin{equation}
%F(u) = \frac{1}{b^2} - u^2 + 2M u^3 . 
%\label{F-Sch-strong}
%\end{equation} 
By using $F(u)$ given by Eq. (\ref{F-Sch}), 
we solve Eq. (\ref{alpha-full}) in an analytic manner. 
The exact expressions involve incomplete elliptic integrals of the first kind. 
When the distances from the lens to the source and the receiver are much larger than the impact parameter of light 
($r_S \gg b, r_R \gg b$) 
but the light ray passes near the photon sphere ($r_0 \sim 3M$), 
Eq. (\ref{alpha-full}) becomes approximately 
\begin{align}
\alpha=
&\frac{2M}{b}\left[\sqrt{1-b^2u_R^2}+\sqrt{1-b^2u_S^2}-2\right]
\notag\\
&+2\log\left(\frac{12(2-\sqrt{3})r_0}{r_0-3M}\right)-\pi
\notag\\
&+O\left(\frac{M^2}{r_R{}^2},\frac{M^2}{r_S{}^2},1-\frac{3M}{r_0}\right) ,
\label{alpha-approx}
\end{align}
where we used a logarithmic term \cite{Iyer} 
in the last term of Eq. (\ref{alpha-full}). 
Here, the dominant terms in $\Psi_R$ and $\Psi_S$ 
cancel with the terms in the integrals. 
As a consequence, $\Psi_R$ and $\Psi_S$ do not appear 
in the approximate expression of Eq. (\ref{alpha-approx}).

As mentioned above, it follows that 
the logarithmic term by the strong gravity 
is free from finite-distance corrections 
such as $\sqrt{1 - (b u_S)^2}$. 
By chance, 
$\delta\alpha$ in the strong deflection limit 
(See Eq. (\ref{alpha-full})) 
is apparently the same as that 
for the weak deflection case 
(See e.g.  Eq. (29) in \cite{Ishihara2017}). 
Therefore, the finite-distance correction 
in the strong deflection limit 
is again 
\begin{align}
\delta\alpha \sim 
O\left(\frac{M b}{r_S{}^2} + \frac{M b}{r_R{}^2}\right) . 
\label{order-alpha}
\end{align}
This is the same expression as that 
for the weak field case (e.g. \cite{Ishihara2016}). 
Namely, the correction is linear in the impact parameter. 
The finite-distance correction in the weak deflection  case (large $b$)
is thus larger than that in the strong deflection one  (small $b$), 
if the other parameters remain the same.

\subsection{Sagittarius A$^{\ast}$}
Next, we briefly mention an astronomical implication 
of the strong deflection. 
One of the most feasible candidates for the strong deflection is 
Sagittarius $^{\ast}$ (Sgr A$^{\ast}$) 
that is located 
at our galactic  center. 
In this case, the receiver distance is much larger 
than the impact parameter of light 
and a source star may live in the bulge of our Galaxy. 

The apparent size of Sgr A$^{\ast}$ is expected to 
be nearly the same as that of the central massive 
object of M87. 
However, the finite-distance correction to Sgr A$^{\ast}$ becomes 
much larger than that to the M87 case, 
because Sgr A$^{\ast}$ is much closer to us than M87. 

For Sgr A$^{\ast}$, Eq. (\ref{order-alpha}) is 
evaluated as 
\begin{align}
\delta\alpha 
&\sim 
\frac{Mb}{r_S{}^2} 
\nonumber\\
&\sim 
10^{-5} \mbox{arcsec.} 
\times 
\left(\frac{M}{4 \times 10^6 M_{\odot}}\right) 
%\times 
\left(\frac{b}{3M}\right)
\left(\frac{0.1 \mbox{pc}}{r_S}\right)^2 , 
\label{alpha-Sgr}
\end{align}
where the central black hole mass is assumed 
as $M \sim 4 \times 10^6 M_{\odot}$ 
and  
we take the limit of strong deflection $b \sim 3M$. 
Rather interestingly, 
this correction as $\sim 10^{-5} \mbox{arcsec.}$ 
will be reachable by the Event Horizon Telescope \cite{EHT} 
and the near-future astronomy.

\begin{figure}
\includegraphics[width=15cm]{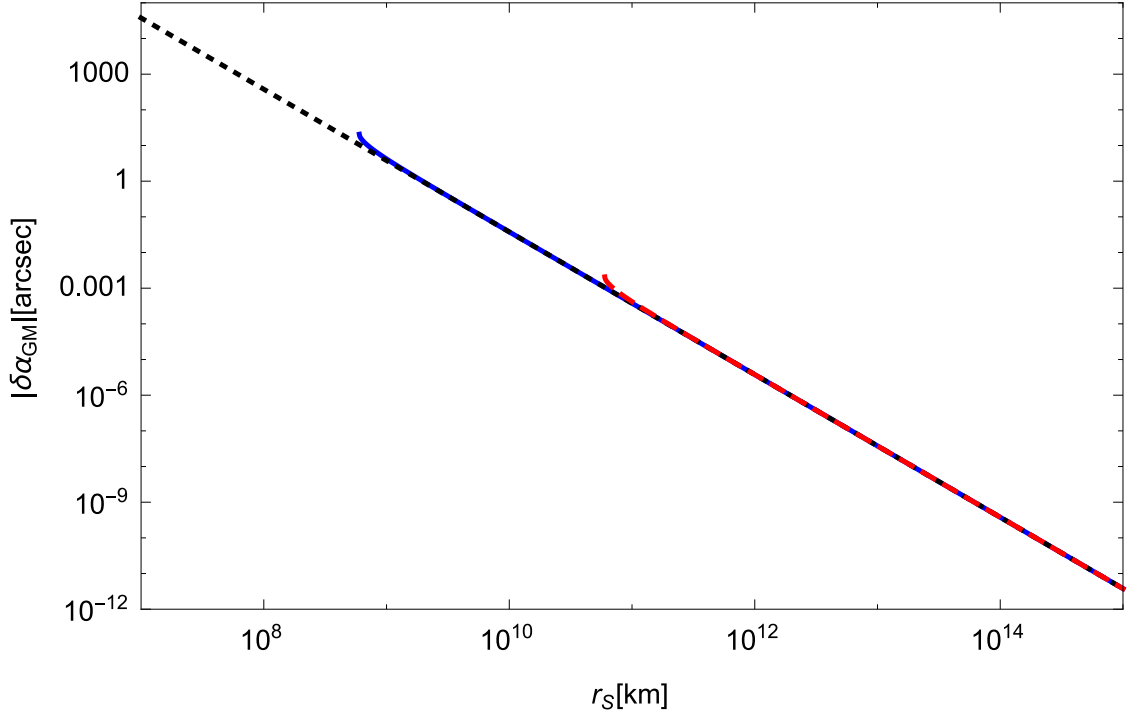}
\caption{ 
The finite-distance correction 
for the Sgr A$^{\ast}$ 
as $\delta\alpha_{GM}$. 
The horizontal axis denotes the source distance $r_S$. 
The vertical one means the finite-distance 
correction to the light deflection. 
The solid line (blue in color) and 
dashed one (red in color) 
mean $b=10^2 M$ and $b=10^4 M$, respectively. 
The dotted curve (yellow in color) denotes 
the leading term of $\delta\alpha_{GM}$ 
given by Eq. (\ref{delta-alpha}). 
These three lines are substantially overlapped 
with each other. 
This implies that 
$\delta\alpha_{GM}$ is weakly dependent 
on the impact parameter $b$. 
}
\label{fig-Sgr}
\end{figure}

See Figure \ref{fig-Sgr} for numerical estimations  
of the finite-distance correction by the source distance. 
This figure and Eq. (\ref{alpha-Sgr}) suggest that 
$\delta\alpha$ 
is $\sim$ ten (or more) micro arcseconds, 
if a source star is sufficiently close to 
Sgr A$^{\ast}$, 
for instance within a tenth of one parsec from 
Sgr A$^{\ast}$. 
For such a case, 
the infinite-distance limit does not hold, 
even though the source is still in the weak field. 
We should take account of finite-distance corrections 
that are discussed in this paper. 

In the strong deflection case, each orbit around 
the black hole will have a slightly different $r_0$, 
thereby producing a number of "ghost" images 
(often called relativistic images). 
In this paper, detailed calculations about it 
for the finite-distance source and receiver are not done. 
It is left for future.

\section{Defining the gravitational deflection angle of light for a 
stationary and axially symmetric spacetime}
\subsection{Optical metric for the stationary,  axisymmetric spacetime}
In this section, a stationary and 
axisymmetric spacetime is considered, 
for which we shall discuss how to define the gravitational deflection  angle of light 
especially with using the Gauss-Bonnet theorem \cite{Ono2017}. 
The line element in this spacetime is \cite{Lewis,LR,Papapetrou}
\begin{align}
ds^2=&g_{\mu\nu}dx^{\mu}dx^{\nu} \notag\\
=&-A(y^1,y^2)dt^2-2H(y^1,y^2)dtd\phi \notag\\
&+F(y^1,y^2) \gamma_{pq}dy^pdy^q + D(y^1,y^2)d\phi^2 . 
\label{ds2-general}
\end{align}
Here, $p, q$ mean $1$ and $2$, 
$\gamma_{pq}$ is a two-dimensional symmetric tensor, 
$\mu, \nu$ take from $0$ to $3$, 
$t$ and $\phi$ coordinates respect 
the Killing vectors. 
We rewrite this metric into a form, 
such that $\gamma_{pq}$ can be diagonal. 
We prefer to use the polar coordinates 
rather than the cylindrical ones, 
because the Kerr metric and the rotating Teo wormhole one are usually expressed in the polar coordinates. 
In the polar coordinates, 
Eq. (\ref{ds2-general}) is rewritten as \cite{Metric}
\begin{align}
ds^2=&-A(r,\theta)dt^2-2H(r,\theta)dtd\phi \notag\\
&+B(r,\theta)dr^2+C(r,\theta)d\theta^2+D(r,\theta)d\phi^2 ,
\label{ds2-axial}
\end{align}
where a local reflection symmetry is assumed 
with respect to the equatorial plane  $\theta=\frac{\pi}{2}$. 
This assumption is expressed as 
\begin{align}
\left.\frac{\partial g_{\mu\nu}}{\partial \theta}\right|_{\theta=\frac{\pi}{2}}=0 .
\label{reflection}
\end{align} 
The functions are $A(r,\theta)>0, B(r,\theta)>0, C(r,\theta)>0, D(r,\theta)>0$ 
and $H(r,\theta)>0$. 
This assumption by Eq. (\ref{reflection}) is needed for the existence of a photon orbit 
on the equatorial plane.
Note that we do not assume the global reflection  symmetry with respect to the equatorial plane. 

The null condition $ds^2 = 0$ is solved for $dt$ as \cite{Civita,AK}
\begin{align}
dt=& \sqrt{\gamma_{ij} dx^i dx^j} +\beta_i dx^i , 
\label{opt} 
\end{align}
where 
$i, j$ denote from $1$ to $3$, 
$\gamma_{ij}$ and $\beta_i$ are defined as  
\begin{align}
\gamma_{ij}dx^idx^j \equiv&
\frac{B(r,\theta)}{A(r,\theta)}dr^2
+\frac{C(r,\theta)}{A(r,\theta)}d\theta^2
+\frac{A(r,\theta)D(r,\theta)+H^2(r,\theta)}{A^2(r,\theta)}d\phi^2 , 
\label{gamma-axial}
\\
\beta_idx^i \equiv& -\frac{H(r,\theta)}{A(r,\theta)} d\phi . 
\label{beta}
\end{align}

This spatial metric $\gamma_{ij} (\neq g_{ij})$ 
is used in order to define the arc length ($\ell$) 
along the photon orbit as 
\begin{align}
d\ell^2 \equiv \gamma_{ij} dx^i dx^j , 
\label{affine}
\end{align}
for which 
we define $\gamma^{ij}$ by  
$\gamma^{ij}\gamma_{jk} = \delta^i_{~k}$. 
$\gamma_{ij}$ defines a 3-dimensional Riemannian space ${}^{(3)}M$, 
where the photon orbit is a spatial curve. 
In the Appendix of Ref. \cite{AK}, they show 
that $\ell$ is an affine parameter of a light ray. 

If the spacetime is static, spherically symmetric 
and asymptotically flat, 
$\beta_i$ is zero and $\gamma_{ij}$ is nothing but the optical metric. 
The photon orbit follows a geodesic 
in a 3-dimensional Riemannian space. 
In this section and after, we refer to $\gamma_{ij}$ 
as the generalized optical metric. 
Note that the metric $\gamma _{ij}$ has been called the Fermat metric and the one-form $\beta _i$ the Fermat one-form by some authors.

We apply Gauss-bonnet theorem to a surface 
(See Figure \ref{fig-GB}). 
The Gauss-Bonnet theorem is expressed as 
\begin{align}
\iint_{{}^{R_{\infty}}_{R}{}^{S_{\infty}}_{S}}KdS 
+\int^{S}_{R} \kappa_g d\ell 
+ \int^{R_{\infty}}_{S_{\infty}} \bar{\kappa}_g d\ell 
+[\Psi_R+(\pi-\Psi_S)+\pi]=2\pi ,
\label{GBT-equa}
\end{align}
where we note that the geodesic curvatures of the path from $S$ to $S_{\infty}$ and 
the path from $R$ to $R_{\infty}$ are both $0$, 
because these paths are geodesic.
$\kappa_g$ is the geodesic curvature of 
the photon orbit and 
$\bar{\kappa}_g$ is the geodesic curvature of the circular arc segment with an infinite radius.

\subsection{Gaussian curvature} 
In this subsection, we examine whether or not 
the rotational part ($\beta_i$) 
of the spacetime makes a contribution to 
the Gaussian curvature. 
The Gaussian curvature on the equatorial plane is expressed 
by using the 2-dimensional Riemann tensor ${}^{(2)}R_{r\phi r\phi}$ as %\cite{Werner2012} 
\begin{align}
K=&\frac{{}^{(2)}R_{r\phi r\phi}}{\det{\gamma^{(2)}_{ij}}} \notag\\
=&\frac{1}{\sqrt{\det{\gamma^{(2)}_{ij}}}}
\left[\frac{\partial}{\partial\phi}
\left(\frac{\sqrt{\det{\gamma^{(2)}_{ij}}}}{\gamma^{(2)}_{rr}}
{{}^{(2)}\Gamma^{\phi}}_{rr}
\right)
-\frac{\partial}{\partial r}
\left(\frac{\sqrt{\det{\gamma^{(2)}_{ij}}}}{\gamma^{(2)}_{rr}}
{{}^{(2)}\Gamma^{\phi}}_{r\phi}\right)\right] , \label{Gaussian}
\end{align} 
where ${}^{(2)}R_{r\phi r\phi}$ and ${}^{(2)}{\Gamma^{\i}}_{jk}$ are 
defined by using the generalized optical metric  $\gamma_{ij}$ on the equatorial plane. 
$\det{\gamma^{(2)}_{ij}}$ is the determinant of the generalized optical metric 
in the equatorial plane. 

$dS$ in Eq.(\ref{GBT-equa}) becomes 
\begin{align}
dS=\sqrt{\det\gamma^{(2)}}drd\phi. 
\label{AElement}
\end{align}
The surface integration of the Gaussian curvature 
in Eq.(\ref{GBT-equa}) is rewritten explicitly as 
\begin{align}
\iint_{{}^{R_{\infty}}_{R}\square^{S_{\infty}}_{S}}KdS =
\int^{\phi_R}_{\phi_S}\int^{\infty}_{r_{OE}} K\sqrt{\det\gamma^{(2)}}drd\phi,
\end{align}
where $r_{OE}$ means the solution of the orbit equation.

\subsection{Geodesic curvature} 
Let us imagine a parameterized curve in a surface. 
Roughly speaking, the geodesic curvature of the parameterized curve is a measure of  
how different the curve is from the geodesic. 
The geodesic curvature of the parameterized curve 
is defined as the surface-tangential component 
of the acceleration (namely the geodesic curvature) of the curve. 
The normal curvature is defined as 
the surface-normal component of the acceleration. 
The normal curvature does not appear 
in the present paper,  
because we consider only the curves 
on the equatorial plane. 

The geodesic curvature in the vector form 
is defined as 
(see e.g. \cite{Math})
\begin{align}
\kappa_g \equiv \bm{T}^{\prime} \cdot \left(\bm{T} \times \bm{N}\right) , 
\label{kappag-vector}
\end{align}
where, for a parameterized curve, 
$\bm{T}$ denotes the unit tangent vector for the curve 
by reparameterizing the curve using its arc length, 
$\bm{T}^{\prime}$ means its derivative with respect 
to the parameter, 
and $\bm{N}$ indicates 
the unit normal vector for the surface. 
The geodesic curvature of a curve vanishes, 
if the curve follows the geodesic. 
This zero is because the acceleration vector $\bm{T}^{\prime}$ vanishes.

\subsection{Photon orbit with the generalized optical metric} \label{tangacce}
In this subsection, we discuss geometrical aspects 
of a photon orbit in terms of the generalized 
optical metric. 
The unit vector tangent to the spatial curve 
is generally expressed as 
\begin{align}
e^i \equiv \frac{dx^i}{d\ell} , 
\end{align} 
where a parameter $\ell$ is defined by Eq.(\ref{affine}). 

The flight time $T$ of a light from the source to the receiver is obtained 
by performing the integral of Eq.(\ref{opt}), 
\begin{align}
T = \int^{t_R}_{t_S} dt 
= \int^{R}_{S} \left(\sqrt{\gamma_{ij} de^i de^j} +\beta_i de^i \right)d\ell . 
\end{align}  
The light ray follows the Fermat's principle, 
namely $\delta T=0$ \cite{PerlickBook}. 
The Lagrangian for a photon can be expressed as 
\begin{align}
\mathcal{L}=\sqrt{\gamma_{ij}e^ie^j}+\beta_ie^i . \label{lagrang}
\end{align}
From this, We obtain   
\begin{align}
\frac{d}{d\ell}\frac{\partial \mathcal{L}}{\partial e^k}=&
\gamma_{ik}{e^i}_{,l}e^l+\gamma_{ik,l}e^ie^l+\beta_{k,i}e^i,\\
\frac{\partial \mathcal{L}}{\partial x^k}=&
\frac12\gamma_{ij,k}e^ie^j+\beta_{i,k}e^i ,
\end{align} 
where we used $\gamma_{ij}e^ie^j=1$ and the comma ($,$) defines the partial derivative. 
The Euler-Lagrange equation is calculated as
\begin{align}
{e^j}_{,l}e^l+\gamma^{kj}\left(\gamma_{ik,l}e^ie^l
-\frac12\gamma_{il,k}e^ie^l\right)
=\gamma^{kj}(\beta_{l,k}-\beta_{k,l})e^l. 
\end{align}
This leads to the equation for the light ray as \cite{AK} 
\begin{align}
\frac{de^i}{d\ell}=
-\gamma^{il}(\gamma_{lj,k}-\frac12\gamma_{jk,l})e^je^k 
+\gamma^{ij}(\beta_{k,j}-\beta_{j,k})e^k .\notag
\end{align}
Therefore, the geodesic equation is equivalent to 
\begin{align}
{e^i}_{|j}e^j=&
\frac{de^i}{d\ell}+{}^{(3)}{\Gamma^{i}}_{jk}e^je^k \notag\\
=&\frac{de^i}{d\ell}+\gamma^{il}(\gamma_{lj,k}-\frac12\gamma_{jk,l})e^je^k \notag\\
=&\gamma^{ij}(\beta_{k,j}-\beta_{j,k})e^k,  \label{EL}
\end{align}
where 
we define $|$ 
as the covariant derivative with respect to $\gamma_{ij}$ . 
${}^{(3)}\Gamma^i_{~jk}$ means the Christoffel symbol 
by $\gamma_{ij}$. 
 
The acceleration vector $a^i$ is defined by 
\begin{align}
a^i \equiv {e^i}_{|j}e^j
=\gamma^{ij}(\beta_{k|j}-\beta_{j|k})e^k 
=\gamma^{ij}(\beta_{k,j}-\beta_{j,k})e^k.
\label{ai}
\end{align}

By using the Levi-Civita symbol $\varepsilon_{ijk}$, 
we express the cross (outer) product 
of $\bm{A}$ and $\bm{B}$ 
in the covariant manner 
\begin{align}
\sqrt{\gamma}\varepsilon_{ijk}A^jB^k=(\bm{A} \times \bm{B})_i . 
\label{cross}
\end{align}
The Levi-Civita tensor 
$\epsilon_{ijk}$ is defined by 
$\epsilon_{ijk} \equiv \sqrt{\gamma}\varepsilon_{ijk}$, 
where  
and $\varepsilon_{ijk}$ is the Levi-Civita symbol 
($\varepsilon_{123} = 1$).
The Levi-Civita tensor $\epsilon_{ijk}$ in a three-dimensional satisfies 
\begin{align}
&\epsilon_{sjk}\epsilon^{slm}
=\sqrt{\gamma} \varepsilon_{sjk} \frac{1}{\sqrt{\gamma}} \varepsilon^{slm}
=\delta_j^l\delta_k^m-\delta_j^m\delta_k^l, \label{levi1} \\
& \epsilon_{sjk} \epsilon^{s}_{~lm}
=\gamma_{jl}\gamma_{km}-\gamma_{jm}\gamma_{kl}. \label{levi2}
\end{align}
By using Eqs.(\ref{cross}), (\ref{levi1}) and (\ref{levi2}), 
Eq.(\ref{ai}) is rewritten as
\begin{align}
a^i=\gamma^{ij}e^k\epsilon_{sjk}(\nabla \times \bm{\beta})^s .
\end{align}

The vector $a^i$ is the spatial vector representing 
the acceleration due to $\beta_i$. 
In particular, $a^i$ is caused in gravitomagnetism \cite{Kopeikin}. 
To be more precise, 
the gravitomagnetic vector has an analogy 
to the Lorentz force in electromagnetism 
$\propto \bm{v} \times (\bm{\nabla} \times \bm{A}_m$), 
in which 
$\bm{A}_m$ denotes the vector potential. 
The vector potential is 
defined as  
$\bm{B}=\nabla \times \bm{A}_m, 
\bm{E}=-\nabla \phi -\frac{\partial \bm{A}_m}{\partial t},$ where $\bm{E}$ and $\bm{B}$  
are the electric and magnetic fields, respectively, 
and the electric potential is $\phi$. 

$\gamma_{ij}$ is not an induced metric but 
the generalized optical metric. 
If $\beta_i$ is non-vanishing, 
the photon orbit 
may be different from a geodesic in ${}^{(3)}M$ with $\gamma_{ij}$, 
even though the light ray in the four-dimensional spacetime follows the null geodesic.

In a stationary and axisymmetric spacetime, 
it is always possible to find out coordinates, 
such that $g_{0i}$ can vanish and $a^i = 0$. 
In this case, the photon orbit 
is considered a spatial geodesic curve in ${}^{(3)}M$. 

We study axisymmetric cases, 
which allow $g_{0i} \neq 0$. 
Therefore,  geodesic curvature $\kappa_g$ does not 
always vanish in the photon orbit 
in the Gauss-Bonnet theorem, 
because 
the geodesic curvature $\kappa_g$ for 
a photon orbit is owing to the 
gravitomagnetic effect. 
This non-vanishing $\kappa_g$ for the photon orbit 
leads to a crucial difference from the SSS case 
\cite{Ishihara2016,Ishihara2017}.

\subsection{Geodesic curvature of a photon orbit} 
Eq. (\ref{kappag-vector}) is rearranged to be 
in the tensor form as 
\begin{align}
\kappa_g = \epsilon_{ijk} N^i a^j e^k , 
\label{kappag-tensor}
\end{align}
where $\vec{T}$ and $\vec{T}^{\prime}$ are 
corresponding to 
$e^k$ and $a^j$, respectively.

In this paper, the acceleration vector of the photon orbit depends on 
$\beta_{i}$. 
Hence, the geodesic curvature for the photon orbit 
also depends on it.
A non-vanishing integral of the geodesic curvature 
along the light ray appears 
in the Gauss-Bonnet theorem Eq. (\ref{localGB}). 

Substituting Eq. (\ref{ai}) into $a^i$ in Eq. (\ref{kappag-tensor}) 
leads to 
\begin{align}
\kappa_g =& \epsilon_{ijk} N^i \gamma^{jl}(\beta_{n|l}-\beta_{l|n})e^n e^k 
\notag\\
% =& \gamma^{ja}N^ie^ke^b \epsilon_{ijk}
% \epsilon_{sab}(\bm{\nabla}\times\bm{\beta})^s\notag\\
=&\gamma^{ja}N^ie^ke^b \epsilon_{ijk}
\epsilon_{sab} \epsilon^{sml}\beta_{l|m}\notag\\
% =&N^ie^ke^b 
% {{\epsilon_{i}}^{a}}_{k}\epsilon_{sab} \epsilon^{sml}
% \beta_{l|m}\notag\\
% =&N_ie_ke^b 
% \epsilon^{aik}\epsilon_{asb} \epsilon^{sml}
% \beta_{l|m}\notag\\
=&N_ie_ke^b 
({\delta^{i}}_{s}{\delta^{k}}_{b}-{\delta^{i}}_{b}{\delta^{k}}_{s}) \epsilon^{sml}
\beta_{l|m}\notag\\
=& - \epsilon^{ijk} N_i \beta_{j|k} , 
\label{kappag-tensor2} 
\end{align}
where we used $\gamma_{ij}e^ie^j = 1$ and $\gamma_{ij}e^iN^j = 0$. 
The unit vector normal to the equatorial plane is 
\begin{align}
N_p = \frac{1}{\sqrt{\gamma^{\theta\theta}}} \delta_p^{\theta} , 
\label{Nv}
\end{align}
where the upward direction is chosen 
without loss of generality. 

For the equatorial plane, we obtain 
\begin{align}
\epsilon^{\theta p q} \beta_{q|p} 
&=-\frac{1}{\sqrt{\gamma}}\beta_{\phi,r} , 
\label{rot-beta}
\end{align}
where 
we use $\epsilon^{\theta r \phi} = - 1/\sqrt{\gamma}$ 
and 
$\beta_{r,\phi} = 0$ because of the axisymmetry. 

By using Eqs. (\ref{Nv}) and (\ref{rot-beta}), 
$\kappa_g$ in Eq. (\ref{kappag-tensor2}) becomes 
\begin{align}
\kappa_g=-\frac{1}{\sqrt{\gamma\gamma^{\theta\theta}}} \beta_{\phi,r} . 
\label{kappag-final}
\end{align}
 
By using Eq.(\ref{affine}), 
the line element in the path integral is 
obtained as 
\begin{align}
d\ell =
\sqrt{\gamma_{rr}\left(\frac{dr}{d\phi}\right)^2+\gamma_{\phi\phi}}d\phi ,
\label{LElementlight}
\end{align}
where $\theta=\pi/2$.

\subsection{Geodesic curvature of a circular arc segment }
In a flat space, 
the geodesic curvature $\kappa$ of 
the circular arc segment of radius $R$ is obtained as
\begin{align}
\kappa=\frac1R.
\end{align}
The geodesic curvature $\bar{\kappa}_g$ of a circular arc segment 
of radius $R_c=R_{\infty}$ is obtained as 
\begin{align}
\bar{\kappa}_g=\frac1R_c, \label{arc}
\end{align}
where the radius $R_c$ is sufficiently larger 
than $r_R$ and $r_S$, and
the circular arc segment is 
in the asymptotically flat region.

Eq.(\ref{affine}) becomes $d\ell^2=dr^2+r^2(d\theta^2+\sin^2\theta d\phi^2)$, 
because we assume an asymptotically flat spacetime. 
Hence, the line element in the path integral of $\bar{\kappa}_g$ 
is obtained as
\begin{align}
d\ell =R_c d\phi , 
\end{align}
where we choose $\theta=\pi/2$ and $r=R_c$ 
for the circular arc segment. 

Therefore, the path integral of $\bar{\kappa}_g$ in Eq.(\ref{GBT-equa}) 
is rewritten as
\begin{align}
\int^{R_{\infty}}_{S_{\infty}} \bar{\kappa}_g d\ell = 
\int^{\phi_R}_{\phi_S}d\phi=\phi_R-\phi_S=\phi_{RS},
\end{align}
where we denote the angular coordinate values of the receiver and the source 
as $\phi_R$ and $\phi_S$, respectively.

\subsection{Impact parameter and light rays} \label{jumpsec}
By using Eq. (\ref{ds2-axial}), 
we study the orbit equation on the equatorial plane. 
The Lagrangian for a photon in the equatorial plane is obtained as 
\begin{align}
\hat{\mathcal{L}}=
-A(r)\dot{t}^2-2H(r)\dot{t}\dot{\phi}+B(r)\dot{r}^2+D(r)\dot{\phi}^2 ,
\end{align}
where the dot denotes the derivative 
with respect to the affine parameter and 
the functions $A(r), B(r), D(r), H(r)$ mean, to be rigorous, 
$A(r,\pi/2), B(r,\pi/2), D(r,\pi/2), H(r,\pi/2)$ respectively.

The metric (or the Lagrangian $\hat{\mathcal{L}}$ 
in the 4-dimensional spacetime) 
is independent of $t$ and $\phi$. 
Therefore, 
\begin{align}
&\frac{d}{d\ell}\frac{\partial\hat{\mathcal{L}}}{\partial \dot{t}}=0, \notag\\
&\frac{d}{d\ell}\frac{\partial\hat{\mathcal{L}}}{\partial \dot{\phi}}=0. \notag
\end{align}
Then, associated with the two Killing vectors $\xi^{\mu}=(1,0,0,0)$ 
and $\bar{\xi}^{\mu}=(0,0,0,1)$,
respectively,
\begin{align}
&\frac{\partial\hat{\mathcal{L}}}{\partial \dot{t}}
=g_{\mu\nu}\xi^{\mu}k^{\nu}, \notag\\
&\frac{\partial\hat{\mathcal{L}}}{\partial \dot{\phi}}
=g_{\mu\nu}\bar{\xi}^{\mu}k^{\nu},
\end{align}
where $k^{\mu}=\frac{dx^{\mu}}{d\ell}$ is the vector 
tangent to the light ray 
in the four-dimensional spacetime. 
There are two constants of motion 
\begin{align}
E&=A(r)\dot{t}+H(r)\dot{\phi} , 
\label{EEE} 
\\
L&=D(r)\dot{\phi}-H(r)\dot{t} , 
\label{L}
\end{align}
where $E$ denotes the energy of the photon and 
$L$ means the angular momentum of the photon. 
The impact parameter of the photon is defined as 
\begin{align}
b &\equiv \frac{L}{E} 
\notag\\ 
&=\frac{-H(r)\dot{t}+D(r)\dot{\phi}}{A(r)\dot{t}+H(r)\dot{\phi}} 
\notag\\
&=\cfrac{-H(r)+D(r)\cfrac{d\phi}{dt}}{A(r)+H(r)\cfrac{d\phi}{dt}} . 
\label{b-axial}
\end{align}

In terms of the impact parameter $b$, 
$\hat{\mathcal{L}}=0$ can be considered 
as the orbit equation 
\begin{align}
\left(\frac{dr}{d\phi}\right)^2
=\frac{A(r)D(r)+H^2(r)}{B(r)}
\frac{D(r)-2H(r)b-A(r)b^2}{\left[H(r)+A(r)b\right]^2} , 
\label{OE}
\end{align}
where we used Eq.(\ref{ds2-axial}). 
By introducing $u \equiv 1/r$, 
we rewrite the orbit equation as 
\begin{align}
\left(\frac{du}{d\phi}\right)^2
= F(u) , 
\label{OE-2}
\end{align}
where $F(u)$ is 
\begin{align}
F(u) 
=\frac{u^4 (AD+H^2) (D-2Hb-Ab^2)}{B (H+Ab)^2} . 
\label{F-axial}
\end{align}

\begin{figure}%[H]
\includegraphics[width=15cm]{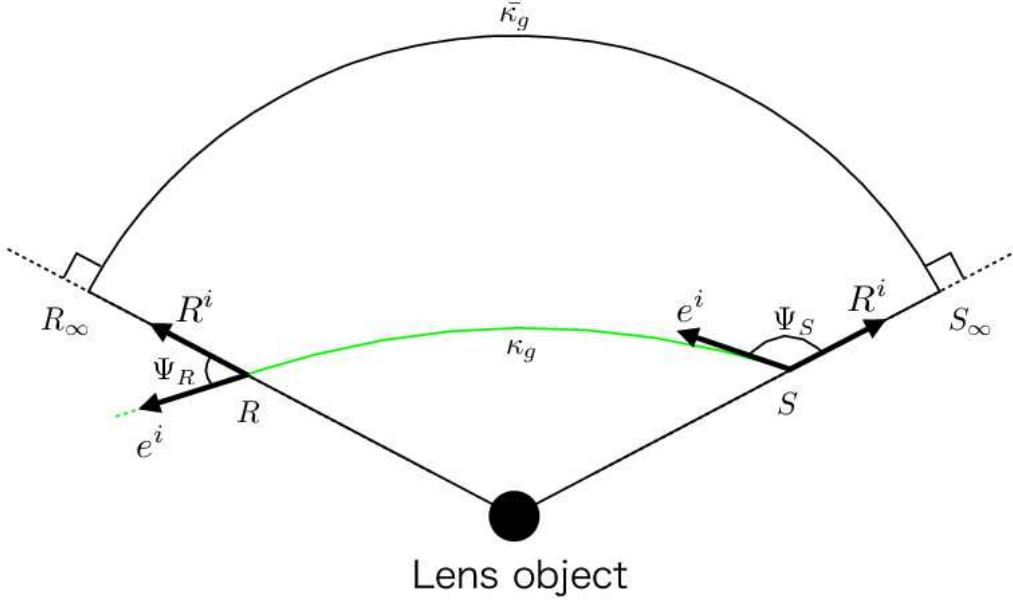}
\caption{
$\Psi_R$ and $\Psi_S$. 
$\Psi_R$ is the angle between the radial direction 
and the light ray at the receiver position. 
$\Psi_S$ is that at the source position. }
\label{Angle}
\end{figure}

We examine the angles 
($\Psi_R, \Psi_S$ in figure \ref{Angle}) 
at the receiver position and the source one. 
The unit vector tangent to the photon orbit in ${}^{(3)}M$ 
is $e^i$. 
Its components on the equatorial plane  
are expressed as 
\begin{align}
e^i=\frac{1}{\xi} \Big(\frac{dr}{d\phi}, 0, 1 \Big) ,
\label{ei}
\end{align}
where $\xi$ satisfies 
\begin{align}
\frac{1}{\xi}=\frac{A(r)[H(r)+A(r)b]}
{A(r)D(r)+H^2(r)} . 
\label{xi}
\end{align}
This can be derived also from $\gamma_{ij} e^i e^j = 1$ 
by using Eq. (\ref{OE}).

In the equatorial plane, the unit radial vector is 
\begin{align}
R^i= \Big(\frac{1}{\sqrt{\gamma_{rr}}}, 0, 0 \Big) , 
\label{R}
\end{align}
where the outgoing direction is chosen 
for a sign convention.

By using the inner product between $e^i$ and $R^i$, therefore, 
we define the angle as 
\begin{align}
\cos \Psi \equiv& 
\gamma_{ij} e^i R^j 
\notag\\
=& \sqrt{\gamma_{rr}}
\frac{A(r)[H(r)+A(r)b]}{A(r)D(r)+H^2(r)}
\frac{dr}{d\phi} , 
\label{cos}
\end{align}
where Eqs. (\ref{ei}), (\ref{xi}) and (\ref{R}) are used. 
This is rewritten as 
\begin{align}
\sin\Psi 
=&\frac{H(r)+A(r)b}
{\sqrt{A(r)D(r)+H^2(r)}} , 
\label{sin}
\end{align}
where Eq. (\ref{OE}) is used. 
We should note that $\sin\Psi$ in Eq. (\ref{sin}) 
is more useful in practical calculations, 
because it needs only the local quantities. 
On the other hand, 
$\cos\Psi$ by Eq. (\ref{cos}) needs the derivative as $dr/d\phi$. 
In addition, the domain of this $\Psi$ is $0 \leq \Psi \leq \pi$ 
and hence $\sin\Psi$ is always positive. 

By substituting $r_R$ and $r_S$ into $r$ of Eq.(\ref{sin}), 
we obtain $\sin\Psi_R$ and $\sin\Psi_S$, respectively. 
We note that the range of the principal value of $y=\arcsin x$ is 
$-\frac{\pi}{2}\leq y \leq \frac{\pi}{2}$ as usual. However, 
the range of $\Psi_R$ ($\Psi_S$) is $0 \leq \Psi_R$($\Psi_S$)$ \leq \pi$.  
By using the usual principal value, 
Eq.(\ref{sin}) for ($\Psi_R$) and ($\Psi_S$) become
\begin{align}
\sin\Psi_R 
=&\frac{H(r_R)+A(r_R)b}
{\sqrt{A(r_R)D(r_R)+H^2(r_R)}}, \label{sinPsiR}\\
\sin(\pi-\Psi_S) 
=&\frac{H(r_S)+A(r_S)b}
{\sqrt{A(r_S)D(r_S)+H^2(r_S)}}, \label{sinPsiS}
\end{align}
respectively, 
because $\Psi_R$ is an acute angle and $\Psi_S$ is an obtuse angle as shown by figure \ref{Angle}.

\subsection{Gravitational deflection light in the axisymmetric case}
We define 
\begin{equation}
\alpha \equiv \Psi_R - \Psi_S + \phi_{RS} ,
\label{alpha-axial}
\end{equation} 
for the equatorial plane in the axisymmetric spacetime. 
This definition apparently depends on 
the angular coordinate $\phi$. 
By using the Gauss-Bonnet theorem Eq. (\ref{localGB}), 
this equation is rearranged as 
\begin{align}
\alpha 
=-\iint_{{}^{\infty}_{R}\square^{\infty}_{S}} K dS 
- \int_{R}^{S} \kappa_g d\ell . 
\label{GB-axial}
\end{align}
Here, $d\ell$ is positive 
when the photon is in the prograde motion, 
whereas it is negative for the retrograde case. 
Eq. (\ref{GB-axial}) means that $\alpha$ is coordinate-invariant for the axisymmetric case. 
Up to now, we do not use any 
equation for gravitational fields. 
Therefore, the above discussion and results 
still stand 
not only in the theory of general relativity  
but also in a general class of metric theories of gravity, 
only if the light ray in the four-dimensional 
spacetime is a null geodesic.

\section{Weak deflection of light in Kerr spacetime} 
\subsection{Kerr spacetime and $\gamma_{ij}$} 
In this section, we focus on the weak deflection of light in the Kerr spacetime 
as an axisymmetric example.
Kerr metric in the Boyer-Lindquist form 
is expressed as 
\begin{align}
ds^2=&-\left(1-\frac{2Mr}{\Sigma}\right)dt^2
-\frac{4aMr\sin^2\theta}{\Sigma}dtd\phi \notag \\
&+\frac{\Sigma}{\Delta}dr^2+\Sigma d\theta^2
+\left(r^2+a^2+\frac{2a^2Mr\sin^2\theta}{\Sigma}\right)\sin^2\theta d\phi^2 , 
\label{Kerr}
\end{align}
where $\Sigma$ and $\Delta$ are defined as 
\begin{align}
\Sigma&\equiv r^2+a^2\cos^2\theta , 
\label{Sigma}
\\
\Delta&\equiv r^2-2Mr+a^2 . 
\end{align}
Using the Gauss Bonnet theorem, 
the deflection angle of light in the Kerr spacetime 
was calculated for the asymptotic source and receiver 
by Werner \cite{Werner2012}. 
However, his method based 
on the osculating metric is limited within the asymptotic case. 
Later, Ono et al. developed a different approach 
using the Gauss-Bonnet theorem that 
enables to calculate the deflection angle 
for the finite distance case in the Kerr spacetime  \cite{Ono2017}.

By using Eqs. (\ref{gamma-axial}) and (\ref{beta}),  
the generalized optical metric $\gamma_{ij}$ 
and the gravitomagnetic term $\beta_i$ 
for the Kerr metric are obtained as 
\begin{align}
\gamma_{ij}dx^idx^j=&
\frac{\Sigma^2}{\Delta(\Sigma-2Mr)}dr^2 
+ \frac{\Sigma^2}{(\Sigma-2Mr)} d\theta^2
\notag\\
&+ \left(r^2+a^2+\frac{2a^2Mr\sin^2\theta}{(\Sigma-2Mr)}\right) 
\frac{\Sigma\sin^2\theta}{(\Sigma-2Mr)} d\phi^2 ,
\label{gamma-Kerr}
%\end{align}
%\begin{align}
\\
\beta_idx^i=&- \frac{2aMr\sin^2\theta}{(\Sigma-2Mr)}d\phi .
\label{beta-Kerr}
\end{align}

Note that $\gamma_{ij}$ has no linear terms 
in the Kerr spin parameter $a$, 
because only $g_{0i}$ in $g_{\mu\nu}$ has 
a linear term in $a$ and 
$g_{0i} \propto H$ contributes to $\gamma_{ij}$ 
through a quadratic term  $g_{0i}g_{0j} \propto H^2$ 
as shown by Eq. (\ref{gamma-axial}). 

In order to calculate the Gaussian curvature $K$ 
of the equatorial plane,
the geodesic curvature $\kappa_g$ of the light ray 
and 
the geodesic curvature $\bar{\kappa_g}$ of the circular arc of an infinite 
radius and the angles $\Psi_R$ and $\Psi_S$, 
we use two approximations 
for the weak field and slow rotation, 
where $M$ and $a$ play a role of 
book-keeping parameters 
though they are dimensional quantities. 

By using Eq.(\ref{OE}), we obtain the orbit equation 
\begin{align}
\left(\frac{dr}{d\phi}\right)^2
=&\frac{b^2
\left\{\frac{a^2}{b^2}+\frac{r}{b} (\frac{r}{b}-\frac{2 M}{b})\right\}^2 
\left\{\frac{a^2}{b^2} (\frac{2 M}{b}+\frac{r}{b})-\frac{4 a M}{b^2}
+ \frac{2 M}{b}-\frac{r}{b}+\frac{r^3}{b^3}\right\}}
{ \frac{r}{b} \{\frac{2 a M}{b^2}+\frac{r}{b}-\frac{2 M}{b}\}^2} \notag\\
=&\frac{r^4}{b^2}-r^2+2Mr-\frac{4r^3}{b^3}aM+\mathcal{O}(a^2), 
\label{KBHdrdpfi}
\end{align}
where the weak-field and slow-rotation approximations 
are used in the last line.
There are no $M$-squared terms in the last line.
The orbit equation becomes 
\begin{align}
\left(\frac{du}{d\phi}\right)^2=F(u)=\frac{1}{b^2}-u^2+2Mu^3-\frac{4u}{b^3}aM
+\mathcal{O}(a^2u^4) . 
\label{dudphiKBH}
\end{align}

We solve iteratively Eq.(\ref{dudphiKBH}). 
In order to find the zeroth order solution, 
we solve the truncated Eq.(\ref{dudphiKBH}) 
\begin{align}
\left(\frac{du}{d\phi}\right)^2=\frac{1}{b^2}-u^2
+\mathcal{O}(Mu^3,aMu^4,a^2u^4) . 
\end{align}
The zeroth order solution for this equation is 
\begin{align}
u=\frac{\sin \phi}{b},
\end{align}
where we use $\left.\frac{du}{d\phi}\right|_{\phi=\pi/2}=0$ as 
the boundary condition. 
This condition means that the closest approach of the photon orbit is expressed as 
$r=r_0=1/u_0, \phi=\pi/2$.      
We assume that the linear-order solution with $M$ is 
$u=\frac{\sin \phi}{b}+ u_1(\phi)M$. 
In order to obtain $u_1(\phi)$, 
we substitute this expression of $u$ 
into the Eq.(\ref{dudphiKBH}) with terms linear in 
$M$
\begin{align}
\left(\frac{du}{d\phi}\right)^2=\frac{1}{b^2}-u^2+2Mu^3
+\mathcal{O}(aMu^4,a^2u^4) . 
\end{align} 
$u_1(\phi)$ is thus obtained as
\begin{align}
u_1(\phi)=\frac{1}{b^2}(1+\cos^2\phi) ,
\end{align}
where we used the boundary condition mentioned above. 
The solution with $a$ is in a form of  
$u=\frac{\sin \phi}{b}+\frac{M}{b^2}(1+\cos^2\phi)+u_2(\phi)a$ .
Since Eq.(\ref{dudphiKBH}) does not include any 
linear term in $a$, 
we find 
$u_2(\phi)=0$.
The solution with $aM$ is 
$u=\frac{\sin \phi}{b}+\frac{M}{b^2}(1+\cos^2\phi)+u_3(\phi)aM$. 
We substitute this solution into Eq.(\ref{dudphiKBH})
\begin{align}
&\frac{a M}{b} \left\{
b^3 \frac{d u_3(\phi)}{d\phi} \cos\phi +b^3 u_3(\phi) \sin\phi
+2\sin\phi\right\}
+\mathcal{O}(a^2u^4)=0. \notag\\
\end{align} 
Hence, $u_3(\phi)$ is obtained as
\begin{align}
u_3(\phi)=-\frac{2}{b^3} .
\end{align}
Bringing the above results together, 
the iterative solution of Eq.(\ref{dudphiKBH}) is expressed as
\begin{align}
u=\frac{\sin \phi}{b}+\frac{M}{b^2}(1+\cos^2\phi)-\frac{2aM}{b^3}
+\mathcal{O}\left(\frac{M^2}{b^3},\frac{a^2}{b^3}\right). 
\label{Sol-OE}
\end{align}

Next, we solve Eq.(\ref{Sol-OE}) for $\phi$. 
We obtain $\phi$ as
\begin{align}
\phi = \begin{cases}
    \arcsin(bu)+ \frac{-2+b^2u^2}{b\sqrt{1-b^2u^2}}M
    +\frac{2aM}{b^2\sqrt{1-b^2u^2}} 
    +\mathcal{O}\left(\frac{M^2}{b^3},\frac{a^2}{b^3}\right)
& (|\phi| < \frac{\pi}{2}) \\
    \pi-\arcsin(bu)-\frac{-2+b^2u^2}{b\sqrt{1-b^2u^2}}M
    -\frac{2aM}{b^2\sqrt{1-b^2u^2}}
    +\mathcal{O}\left(\frac{M^2}{b^3},\frac{a^2}{b^3}\right)
& ( \frac{\pi}{2} < |\phi|)
  \end{cases} , \label{KBHphi}
\end{align} 
where we can choose the domain of $\phi$ to be $-\pi\le \phi< \pi$ 
without loss of generality.
In the following, the range of the angular coordinate value $\phi_S$ 
at the source point is $-\frac{\pi}{2} \le \phi_S < \frac{\pi}{2}$ and 
the range of the angular coordinate value $\phi_R$ 
at the receiver point is $|\phi_R| > \frac{\pi}{2}$.
We find $|bu| < 1$, because the square root in Eq.(\ref{KBHphi}) must be real 
and nonzero, and the value of $b$ and $u$ are positive.
Therefore, $bu$ satisfies $0< bu <1$ in our calculation.

\subsection{Gaussian curvature on the equatorial plane}
Let us explain how to compute 
the Gaussian curvature by using Eq.(\ref{Gaussian}).
In the Kerr case, it becomes 
\begin{align}
K=&\frac{M \left(-6 r \left(a^2+M^2\right)+6 a^2 M+7 M r^2-2 r^3\right)}
{r^5 (r-2 M)} \notag\\
=&-\frac{2M}{r^3} + \mathcal{O}\left(\frac{M^2}{r^4},\frac{a^2M}{r^5}\right),
\label{KBH-K}
\end{align}
where the weak-field and slow-rotation approximations 
are used in the last line.

Next, we discuss the area element on the equatorial plane by using Eq.(\ref{AElement}).
In the Kerr case, the area element of the equatorial plane is expressed as 
\begin{align}
dS=[r+3M+\mathcal{O}(M^2/r)]drd\phi. \label{AElementKBH}
\end{align}
 
By using Eqs.(\ref{KBH-K}) and (\ref{AElementKBH}),  
the surface integral of the Gaussian curvature in Eq.(\ref{GB-axial}) 
is performed as
\begin{align}
-\iint_{{}^{R_{\infty}}_{R}\square^{S_{\infty}}_{S}}KdS 
=&
\int_{\phi_S}^{\phi_R} \int_{\infty}^{r_{OE}} (-\frac{2M}{r^3}r) dr d\phi
+\mathcal{O}\left(\frac{M^2}{b^2},\frac{aM^2}{b^3},\frac{a^2M}{b^3}\right) 
\notag\\
=&2M 
\int_{\phi_S}^{\phi_R} \int_{0}^{\frac{1}{b}\sin\phi+\frac{M}{b^2}(1+\cos^2\phi)
-\frac{2aM}{b^3}} du d\phi 
+\mathcal{O}\left(\frac{M^2}{b^2},\frac{aM^2}{b^3},\frac{a^2M}{b^3}\right)
\notag\\
%=&2M\int_{\phi_S}^{\phi_R}
%\Big[u\Big]^{\frac{1}{b}\sin\phi+\frac{M}{b^2}(1+\cos^2\phi)
%-\frac{2aM}{b^3}}_{0} d\phi 
%+\mathcal{O}\left(\frac{M^2}{b^2},\frac{aM^2}{b^3},\frac{a^2%M}{b^3}\right)
%\notag\\
=&2M\int_{\phi_S}^{\phi_R}
\Big[\frac{1}{b}\sin\phi\Big] d\phi
+\mathcal{O}\left(\frac{M^2}{b^2},\frac{aM^2}{b^3},\frac{a^2M}{b^3}\right)
\notag\\
%=&\frac{2M}{b}\Big[\cos\phi\Big]^{\phi_S}_{\phi_R} 
%+\mathcal{O}\left(\frac{M^2}{b^2},\frac{aM^2}{b^3},\frac{a^2%M}{b^3}\right)
%\notag\\
=&\frac{2M}{b}\Big[\cos\phi_S-\cos\phi_R\Big]
+\mathcal{O}\left(\frac{M^2}{b^2},\frac{aM^2}{b^3},\frac{a^2M}{b^3}\right)
\notag\\
=&\frac{2M}{b}\Big[\sqrt{1-b^2{u_S}^2}+\sqrt{1-b^2{u_R}^2}\Big]
+\mathcal{O}\left(\frac{M^2}{b^2},\frac{aM^2}{b^3},\frac{a^2M}{b^3}\right) ,
\label{INTK}
\end{align}
where $r_{OE}$ in the first line is the solution of Eq.(\ref{KBHdrdpfi}), 
we transform the integral variable as $r=1/u$ in the second line, 
we used $\cos\phi_S=\sqrt{1-b^2{u_S}^2}+\mathcal{O}(M/b)$ and 
$\cos\phi_R=-\sqrt{1-b^2{u_R}^2}+\mathcal{O}(M/b)$ from Eq.(\ref{KBHphi}) 
in the last line.

\subsection{Path integral of $\kappa_g$}
Substituting Eq. (\ref{beta-Kerr}) 
into $\beta_i$ in Eq. (\ref{kappag-final}) 
leads to 
\begin{align}
\kappa_g
=&
-\frac{2aM}{r^2(r-2M)}
\left(
\cfrac{1-\cfrac{2M}{r}+\cfrac{a^2}{r^2}}
{1+\cfrac{a^2}{r^2}+\cfrac{2a^2M}{r^3}}
\right)^{1/2}
 \notag\\
=&-\frac{2aM}{r^3} + 
\mathcal{O}\left(\frac{aM^2}{r^4}\right) , 
\label{kappa-Kerr}
\end{align}
where the weak-field and slow-rotation approximations 
are used in the last line. 
We stress that the terms of $a^nM$ $(n \geq 2)$ do not exist in this expression. 

The line element for the path integral by Eq.(\ref{LElementlight}) becomes 
\begin{align}
d\ell = \left[\frac{b}{\sin^2\phi}+\mathcal{O}(M)\right]d\phi ,
\label{KBHLElementlight}
\end{align}
where Eq.(\ref{Sol-OE}) was used for a relation between $r$ and $\phi$.

By using (\ref{kappa-Kerr}) and Eqs.(\ref{KBHLElementlight}),
the path integral of $\kappa_g$ in Eq.(\ref{GB-axial}) is performed as 
\begin{align}
-\int^{S}_{R}\kappa_gd\ell
=&-\int^{R}_{S}\frac{2aM}{r^3}d\ell +\mathcal{O}\left(\frac{aM^2}{r^4}\right)
\notag\\
=&-\frac{2aM}{b^2}\int^{\phi_R}_{\phi_S} \sin\phi d\phi 
+\mathcal{O}\left(\frac{aM^2}{r^4}\right)\notag\\
%=&-\frac{2aM}{b^2}[\cos\phi_S-\cos\phi_R] 
%+\mathcal{O}\left(\frac{aM^2}{b^3}\right)\notag\\
=&-\frac{2aM}{b^2}[\sqrt{1-b^2{u_R}^2}+\sqrt{1-b^2{u_S}^2}]
+\mathcal{O}\left(\frac{aM^2}{b^3}\right) . 
\label{int-kappag}
\end{align}
Here, we assumed $d\ell > 0$, 
such that the orbital angular momentum 
can be parallel with the spin of the black hole 
and 
we used a linear approximation of the photon orbit 
as $1/r=u= \sin\phi/b + \mathcal{O}(M/b^2, aM/b^3)$ from Eq.(\ref{Sol-OE}). 
In the retrograde case, 
$d\ell$ becomes negative 
and the magnitude of the above path integral 
thus remains the same 
but the sign of the integral is opposite.

\subsection{$\phi_{RS}$ part}
The displacement of the angular coordinate $\phi$ 
in Eq.(\ref{alpha-axial}) is computed as 
\begin{align}
\phi_{RS} 
=& 
\int^R_S d\phi 
\notag\\
=&
2\int^{u_0}_{0}\frac{1}{\sqrt{F(u)}}du 
+\int^{0}_{u_S}\frac{1}{\sqrt{F(u)}}du +\int^{0}_{u_R}\frac{1}{\sqrt{F(u)}}du , 
\label{phi-Kerr}
\end{align}
where the orbit equation by Eq.(\ref{OE-2}) was 
made use of. 
We substitute Eq.(\ref{dudphiKBH}) into $F(u)$ in Eq.(\ref{phi-Kerr}) to obtain 
\begin{align}
\phi_{RS} 
=&
\int^{u_0}_{u_S}\left(\frac{1}{\sqrt{{u_0}^2-u^2}}
+M\frac{{u_0}^3-u^3}{({u_0}^2-u^2)^{3/2}}
-2aM\frac{{u_0}^3(u_0-u)}{({u_0}^2-u^2)^{3/2}}\right) du 
\notag\\
&
+\int^{u_0}_{u_R}\left(\frac{1}{\sqrt{{u_0}^2-u^2}}
+M\frac{{u_0}^3-u^3}{({u_0}^2-u^2)^{3/2}}
-2aM\frac{{u_0}^3(u_0-u)}{({u_0}^2-u^2)^{3/2}}\right) du 
\notag\\ 
&+\mathcal{O}(M^2{u_0}^2, a^2{u_0}^2) 
\notag\\
=&\left(\frac{\pi}{2}-\arcsin\Big(\frac{u_S}{u_0}\Big)
+M\frac{(2u_0+u_S)\sqrt{{u_0}^2-
u_S^2}}
{u_0+u_S}
-2aM\frac{{u_0}^3\sqrt{{u_0}^2-
u_S^2}}
{{u_0}^2+u_0u_S}\right) \notag\\
&+\left(\frac{\pi}{2}-\arcsin\Big(\frac{u_R}{u_0}\Big)
+M\frac{(2u_0+u_R)\sqrt{{u_0}^2-{u_R}^2}}{u_0+u_R}
-2aM\frac{{u_0}^3\sqrt{{u_0}^2-{u_R}^2}}{{u_0}^2+u_0u_R}\right) 
\notag\\
&+\mathcal{O}\left(M^2 u_0^2, a^2 u_0^2\right) , 
\label{phi-Kerr1}
\end{align}
where the prograde case is assumed. 
In the retrograde motion, the sign of the linear term 
in $a$ is opposite. 
In Eq.(\ref{phi-Kerr1}), the impact parameter $b$ is rewritten in terms of 
the closest approach $u_0$ for the integration 
from $u_S$(or $u_R$) to $u_0$. 
Namely, Eq.(\ref{dudphiKBH}) tells us 
the relation between the impact parameter $b$ and  
the inverse of the closest approach $u_0$ as 
$b = u_0^{-1} + M - 2 aM u_0 +\mathcal{O}(M^2u_0, a^2u_0)$ 
in the weak field and slow rotation approximations. 
By making use of this relation, 
Eq. (\ref{phi-Kerr1}) is rearranged as  
\begin{align}
\phi_{RS} =& \pi-\arcsin(bu_S)-\arcsin(bu_R)
+\frac{M(2-b^2{u_S}^2)}{b\sqrt{1-b^2{u_S}^2}}
+\frac{M(2-b^2{u_R}^2)}{b\sqrt{1-b^2{u_R}^2}} \notag\\
&-\frac{2aM}{b^2}\Big[\frac{1}{\sqrt{1-b^2{u_S}^2}}
+\frac{1}{\sqrt{1-b^2{u_R}^2}}\Big] 
+\mathcal{O}\left(M^2/b^2, a^2/b^2\right). 
\label{phi-Kerr2}
\end{align}
The first line of this equation recovers Eq. (32) of Reference \cite{Ishihara2016}. 

\subsection{$\Psi$ parts} 
In the Kerr spacetime by Eq.(\ref{Kerr}),  Eq.(\ref{sinPsiR}) is 
\begin{align}
\sin\Psi_R
=&\frac{b}{r_R}\times \cfrac{1-\cfrac{2M}{r_R}+\cfrac{2aM}{br_R}}
{\sqrt{1-\cfrac{2M}{r_R}+\cfrac{a^2}{{r_R}^2}}} ,\notag\\
=&\frac{b}{r_R} 
\left(1-\frac{M}{r_R}+\frac{2aM}{br_R}\right) 
+\mathcal{O}
\left(\frac{M^2}{{r_R}^2}, \frac{a^2}{{r_R}^2}, \frac{aM^2}{{r_R}^3} \right) 
\notag\\
=&bu_R 
\left(1-M u_R+\frac{2aMu_R}{b}\right) 
+\mathcal{O}\left(M^2{u_R}^2, a^2{u_R}^2, aM^2{u_R}^3 \right),
\label{KBHsinPR} 
\end{align}
and Eq.(\ref{sinPsiS}) is calculated as 
\begin{align}
\sin(\pi-\Psi_S)=
bu_S \left(1-M u_S+\frac{2aMu_S}{b}\right) 
+\mathcal{O}\left(M^2{u_S}^2, a^2{u_S}^2, aM^2{u_S}^3 \right), 
\label{KBHsinPS} 
\end{align}
where $r_R=1/u_R, r_S=1/u_S$ and 
we used the weak-field and slow-rotation approximations.  
By combining 
Eqs.(\ref{KBHsinPR}) and (\ref{KBHsinPS}), 
we obtain $\Psi_R$ and $\Psi_S$ as 
\begin{align}
\Psi_R=&\arcsin\left[bu_R\left(1-M u_R+\frac{2aMu_R}{b}\right)\right]
+\mathcal{O}\left(M^2{u_R}^2, a^2{u_R}^2, aM^2{u_R}^3 \right) \notag\\
=&\arcsin(bu_R)-\frac{Mb{u_R}^2}{\sqrt{1-b^2{u_R}^2}}
+\frac{2aM{u_R}^2}{\sqrt{1-b^2{u_R}^2}} 
+\mathcal{O}\left(M^2{u_R}^2, a^2{u_R}^2, aM^2{u_R}^3 \right), \notag\\
\pi-\Psi_S=&\arcsin(bu_S)-\frac{Mb{u_S}^2}{\sqrt{1-b^2{u_S}^2}}
+\frac{2aM{u_S}^2}{\sqrt{1-b^2{u_S}^2}} 
+\mathcal{O}\left(M^2{u_S}^2, a^2{u_S}^2, aM^2{u_S}^3 \right).\notag\\
\end{align}
By combining these relations, 
we obtain the $\Psi$ part in Eq.(\ref{alpha-axial}) 
as 
\begin{align}
\Psi_R-\Psi_S
=& \arcsin(bu_R)+\arcsin(bu_S)-\pi -\frac{Mb{u_R}^2}{\sqrt{1-b^2{u_R}^2}}
- \frac{Mb{u_S}^2}{\sqrt{1-b^2{u_S}^2}} 
\notag\\
& +\frac{2aM{u_R}^2}{\sqrt{1-b^2{u_R}^2}} 
+ \frac{2aM{u_S}^2}{\sqrt{1-b^2{u_S}^2}} 
+ \mathcal{O}\left(M^2 u_R^2, M^2 u_S^2, 
a^2 u_R^2, a^2 u_S^2, 
aM^2 u_R^3, aM^2 u_S^3
\right) . 
\label{Psi-Kerr}
\end{align}

\subsection{Deflection of light in Kerr spacetime}
On the equatorial plane in the Kerr spacetime, 
the deflection angle of light is described by  Eq.(\ref{alpha-axial}) and 
Eq.(\ref{GB-axial}).
Let us examine whether the two results agree with each other. 

First, we substitute Eqs. (\ref{phi-Kerr2}) and (\ref{Psi-Kerr}) 
into Eq. (\ref{alpha-axial}). 
We obtain the deflection angle of light as
\begin{align}
\alpha_{prog}=&
\arcsin(bu_R)+\arcsin(bu_S)-\pi 
-\frac{Mb{u_R}^2}{\sqrt{1-b^2{u_R}^2}}
- \frac{Mb{u_S}^2}{\sqrt{1-b^2{u_S}^2}} \notag\\
& +\frac{2aM{u_R}^2}{\sqrt{1-b^2{u_R}^2}} 
+ \frac{2aM{u_S}^2}{\sqrt{1-b^2{u_S}^2}} \notag\\
&+\pi-\arcsin(bu_S)-\arcsin(bu_R)
+\frac{M(2-b^2{u_S}^2)}{b\sqrt{1-b^2{u_S}^2}}
+\frac{M(2-b^2{u_R}^2)}{b\sqrt{1-b^2{u_R}^2}} \notag\\
&-\frac{2aM}{b^2}\Big[\frac{1}{\sqrt{1-b^2{u_S}^2}}
+\frac{1}{\sqrt{1-b^2{u_R}^2}}\Big]
+\mathcal{O}\left(\frac{M^2}{b^2}\right)\notag\\
=&\frac{2M}{b}
\left(\sqrt{1-b^2{u_R}^2}+\sqrt{1-b^2{u_S}^2}\right)
\notag\\
&-\frac{2aM}{b^2}
\left(\sqrt{1-b^2{u_R}^2}+\sqrt{1-b^2{u_S}^2}\right)
+ \mathcal{O}\left(\frac{M^2}{b^2}\right) , 
\label{alpha+}
\end{align}
where the prograde orbit of light is assumed. 
For the retrograde motion, we obtain 
\begin{align}
\alpha_{retro}
=&\frac{2M}{b}
\left(\sqrt{1-b^2{u_R}^2}+\sqrt{1-b^2{u_S}^2}\right)
\notag\\
&+\frac{2aM}{b^2}
\left(\sqrt{1-b^2{u_R}^2}+\sqrt{1-b^2{u_S}^2}\right)
+ \mathcal{O}\left(\frac{M^2}{b^2}\right) . 
\label{alpha-}
\end{align}

Next, we substitute Eqs.(\ref{INTK}) and (\ref{int-kappag}) 
into Eq.(\ref{GB-axial}). 
Then, we obtain the deflection angle of light 
in the prograde motion as
\begin{align}
\alpha_{prog}
=&
\frac{2M}{b}
\left(\sqrt{1-b^2{u_R}^2}+\sqrt{1-b^2{u_S}^2}\right)
\notag\\
&-\frac{2aM}{b^2}
\left(\sqrt{1-b^2{u_R}^2}+\sqrt{1-b^2{u_S}^2}\right)
+ \mathcal{O}\left(\frac{M^2}{b^2}\right) , 
\label{Alpha+}
\end{align}
and the deflection angle for the retrograde case as  
\begin{align}
\alpha_{retro}
=&\frac{2M}{b}
\left(\sqrt{1-b^2{u_R}^2}+\sqrt{1-b^2{u_S}^2}\right)
\notag\\
&+\frac{2aM}{b^2}
\left(\sqrt{1-b^2{u_R}^2}+\sqrt{1-b^2{u_S}^2}\right)
+ \mathcal{O}\left(\frac{M^2}{b^2}\right) . 
\label{Alpha-}
\end{align}
Note that $a^2$ terms in the deflection angle 
in Eq.(\ref{alpha-axial}) cancel out thanks to   Eq.(\ref{GB-axial}). 
%$\sqrt{1-b^2{u_R}^2}$ and $\sqrt{1-b^2{u_S}^2}$ parts 
%in Eqs.(\ref{Alpha+}) 
%and (\ref{Alpha-}) are real numbers so as to ensure %that $bu_R,bu_S<1$ which  
%comes from the note after Eq.(\ref{KBHphi}). 

Here, we consider the limit 
as $u_R \to 0$ and $u_S \to 0$. 
In this limit, we get 
\begin{align}
\alpha_{\infty\, prog} \to 
&\frac{4M}{b}-\frac{4aM}{b^2} 
+ O\left(\frac{M^2}{b^2}\right) , 
\\
\alpha_{\infty\, retro}\to 
&\frac{4M}{b}+\frac{4aM}{b^2}
+ O\left(\frac{M^2}{b^2}\right) . 
\end{align}
This shows that Eqs. (\ref{alpha+}) and (\ref{alpha-}) 
agree with the asymptotic deflection angles that are known in earlier works
\cite{Ch,ES,Ibanez,IH,Kerr-bending}.

If we wish to consider the deflection angle of light 
in a case where the receiver point is closer to the source point 
than the closest approach point, 
Eqs.(\ref{alpha+}) and (\ref{alpha-}) become 
\begin{align}
\alpha_{prog}
=&
\frac{2M}{b}
\left(\sqrt{1-b^2{u_S}^2}-\sqrt{1-b^2{u_R}^2}\right)
\notag\\
&-\frac{2aM}{b^2}
\left(\sqrt{1-b^2{u_S}^2}-\sqrt{1-b^2{u_R}^2}\right)
+ \mathcal{O}\left(\frac{M^2}{b^2}\right) , \notag\\
\alpha_{retro}
=&\frac{2M}{b}
\left(\sqrt{1-b^2{u_S}^2}-\sqrt{1-b^2{u_R}^2}\right)
\notag\\
&+\frac{2aM}{b^2}
\left(\sqrt{1-b^2{u_S}^2}-\sqrt{1-b^2{u_R}^2}\right)
+ \mathcal{O}\left(\frac{M^2}{b^2}\right) .\notag
\end{align}
If we wish to consider the deflection angle of light 
in such a case that the source point is closer to the receiver 
than the closest approach point, 
Eqs.(\ref{alpha+}) and (\ref{alpha-}) become 
\begin{align}
\alpha_{prog}
=&
\frac{2M}{b}
\left(\sqrt{1-b^2{u_R}^2}-\sqrt{1-b^2{u_S}^2}\right)
\notag\\
&-\frac{2aM}{b^2}
\left(\sqrt{1-b^2{u_R}^2}-\sqrt{1-b^2{u_S}^2}\right)
+ \mathcal{O}\left(\frac{M^2}{b^2}\right) , \notag\\
\alpha_{retro}
=&\frac{2M}{b}
\left(\sqrt{1-b^2{u_R}^2}-\sqrt{1-b^2{u_S}^2}\right)
\notag\\
&+\frac{2aM}{b^2}
\left(\sqrt{1-b^2{u_R}^2}-\sqrt{1-b^2{u_S}^2}\right)
+ \mathcal{O}\left(\frac{M^2}{b^2}\right) .\notag
\end{align}

\subsection{Finite-distance corrections} 
In the previous subsections so far 
we discussed an effect of the spin of the lens object 
to the deflection of light. 
In particular, we do not require that the receiver and the source are located at the infinity. 
The finite-distance correction to the deflection angle of light is 
defined as $\delta\alpha$. 
This is the difference
between the asymptotic deflection angle 
$\alpha_{\infty}$ 
and the deflection angle for the finite distance case. 
Namely,  
\begin{align}
\delta\alpha \equiv \alpha-\alpha_{\infty} . 
\end{align}
Equations (\ref{alpha+}) and (\ref{alpha-})
tell us the magnitude of the finite-distance correction to 
the gravitomagnetic bending angle due to the spin. 
The result is  
\begin{align}
\left| \delta\alpha_{GM} \right|
\sim& 
O\left(\frac{aM}{r_S^2} + \frac{aM}{r_R^2}\right) 
\notag\\
\sim&O\left(\frac{J}{r_S^2} + \frac{J}{r_R^2}\right) ,  
\label{delta-alpha-axial}
\end{align}
where $bu_R, bu_S<1$ is assumed, 
$J \equiv aM$ denotes 
the spin angular momentum of the lens 
and the subscript $GM$ means the gravitomagnetic part. 
We introduce the dimensionless spin parameter as 
$s \equiv a/M$. 
Hence, Eq. (\ref{delta-alpha-axial}) 
is rearranged as 
\begin{align}
\left| \delta\alpha_{GM} \right| 
\sim O\left( s\left(\frac{M}{r_S}\right)^2 
+ s\left(\frac{M}{r_R}\right)^2 \right) . 
\label{delta-alpha2-axial}
\end{align}
This implies that $\delta\alpha_{GM}$ 
is of the same order as 
the second post-Newtonian effect 
(with the dimensionless spin parameter). 

The second-order Schwarzschild contribution 
to $\alpha$ 
is $15\pi M^2/4 b^2$. 
This contribution can be obtained also 
by using the present method, 
especially by using a relation 
between $b$ and $r_0$ in $M^2$ 
in calculating $\phi_{RS}$. 
Appendix A provides detailed calculations 
at the second order of $M$ and $a$. 
We explain detailed calculations 
for the integrals of $K$ and  $\kappa_g$ 
in the present formulation.  
Note that $\delta\alpha_{GM}$ in the above approximations 
is free from the impact parameter $b$. 
We can see this fact from Figure \ref{fig-Sun} and Figure \ref{fig-Sgr} below. 

%\begin{figure}
%\includegraphics[width=10cm]{fig-3.eps}
%\caption{ 
%$\delta\alpha_{GM}$ for the Sun. 
%The vertical axis denotes the finite-distance correction 
%to the gravitomagnetic deflection angle of light 
%and the horizontal axis denotes the receiver distance $r_R$. 
%The solid curve (blue in color) and dashed one (red in color) 
%correspond to $b=R_{\odot}$ and $b=10 R_{\odot}$, respectively. 
%The dotted line (black in color) denotes the leading term 
%of $\delta\alpha_{GM}$ given by Eq. (\ref{delta-alpha}). 
%The overlap between these curves suggest that 
%the dependence of $\delta\alpha_{GM}$ on the impact parameter $b$ 
%is very weak. }
%\label{fig-Sun}
%\end{figure}

\subsection{Possible astronomical applications}
What are possible astronomical applications? 
As a first example, we consider the Sun, 
in which its higher multipole moments are ignored 
for its simplicity. 
Its spin angular momentum denoted as $J_{\odot}$ is 
$\sim 2\times 10^{41} \,\mbox{m}^2\,\mbox{kg}\,\mbox{s}^{-1}$ 
\cite{Sun}. 
This means 
$G J_{\odot} c^{-2} \sim 5 \times 10^5 \,\mbox{m}^2$, 
for which the dimensionless spin parameter becomes  
$s_{\odot} \sim 10^{-1}$. 

Here, our assumption is that a receiver on the Earth observes 
the light deflected by the Sun, while 
the distant source is safely 
in the asymptotic region. 
For the light ray passing near the Sun, 
Eq. (\ref{delta-alpha2-axial}) allows us to 
make an order-of-magnitude estimation of 
the finite-distance correction. 
The result is  
\begin{align}
\left| \delta\alpha_{GM} \right| 
&\sim 
O \left(\frac{J}{r_R^2}\right) 
\nonumber\\
&\sim 
10^{-12} \mbox{arcsec.} 
\times 
\left(\frac{J}{J_{\odot}}\right) 
\left(\frac{1 \mbox{AU}}{r_R}\right)^2 , 
\label{alpha-Sun}
\end{align}
where $4M_{\odot}/R_{\odot} \sim 1.75 \,\mbox{arcsec.} 
\sim 10^{-5} \,\mbox{rad.}$, 
$M_{odot}$ means the solar mass
and $R_{\odot}$ denotes the solar radius. 
This correction is nearly a pico-arcsecond. 
Therefore, the correction is 
beyond the reach of  
present and near-future technology 
\cite{Gaia, JASMINE}. 

\begin{figure}%[H]
\includegraphics[width=15cm]{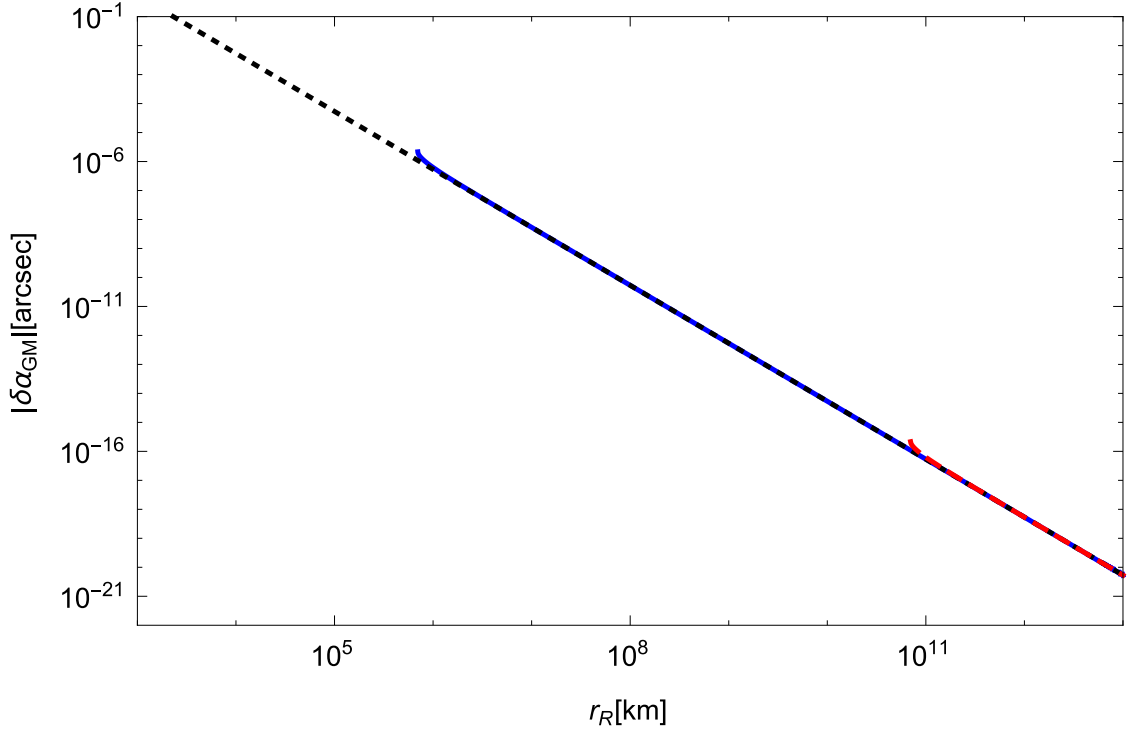}
\caption{ 
$\delta\alpha_{GM}$ for the Sun. 
The horizontal axis is  
the distance of the receiver distance $r_R$. 
The vertical axis means 
the finite-distance 
correction due to the gravitomagnetic deflection 
angle of light. 
The solid curve (blue in color) and dashed one (red in color) 
denote 
$b=R_{\odot}$ and $b=10^5 R_{\odot}$, respectively. 
The dotted line (black in color) corresponds to 
the leading term 
in $\delta\alpha_{GM}$ given by Eq. (\ref{delta-alpha-axial}). 
These three curves are overlapped. 
This implies  
the very weak dependence of $\delta\alpha_{GM}$ 
on $b$. }
\label{fig-Sun}
\end{figure}

Figure \ref{fig-Sun} shows the finite-distance correction to the light deflection. 
Our numerical calculations are consistent with 
the above order-of-magnitude estimation. 
This figure shows also  
the very weak dependence of $\delta\alpha$ 
on $b$. 

\begin{figure}%[H]
\includegraphics[width=15cm]{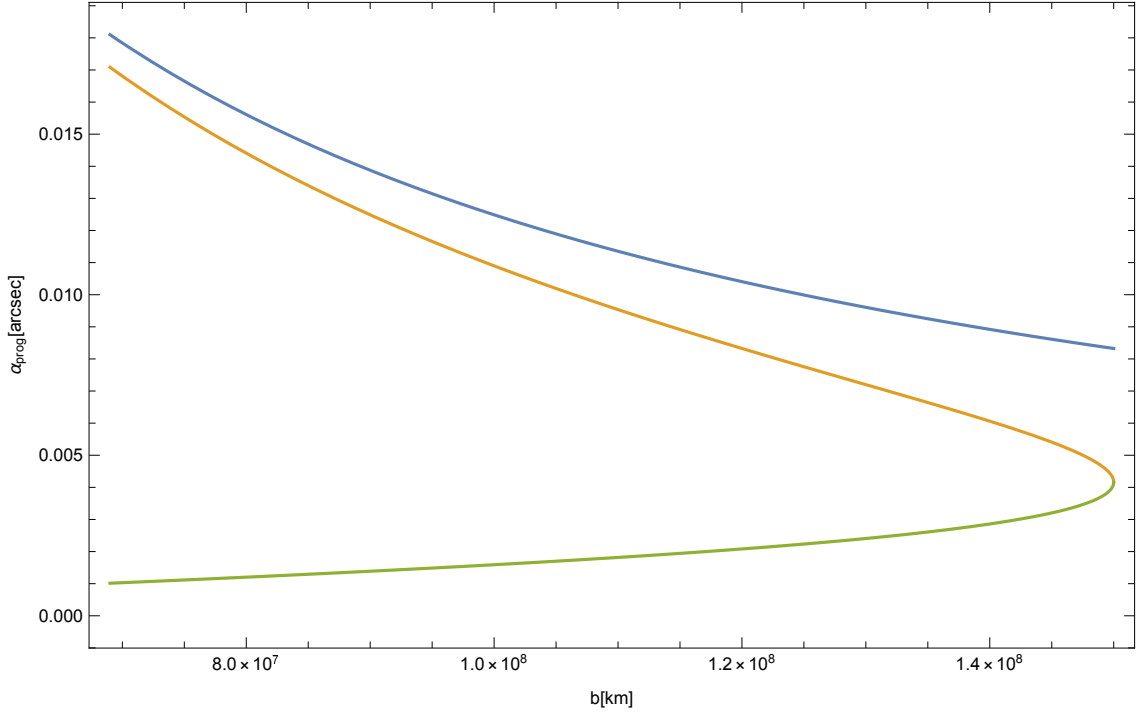}
\caption{ 
$\alpha$ in the prograde motion.
The horizontal axis is the impact parameter for a photon orbit.  
The vertical axis means the deflection angle of light.
The blue curve is the asymptotic deflection angle by a Kerr black hole.
The orange curve means the deflection angle with finite-corrections by a Kerr black hole.
The green curve shows the difference between the asymptotic bending angle 
and the deflection angle with finite-corrections by a Kerr black hole.}
\label{difference}
\end{figure}

\begin{figure}%[H]
\includegraphics[width=15cm]{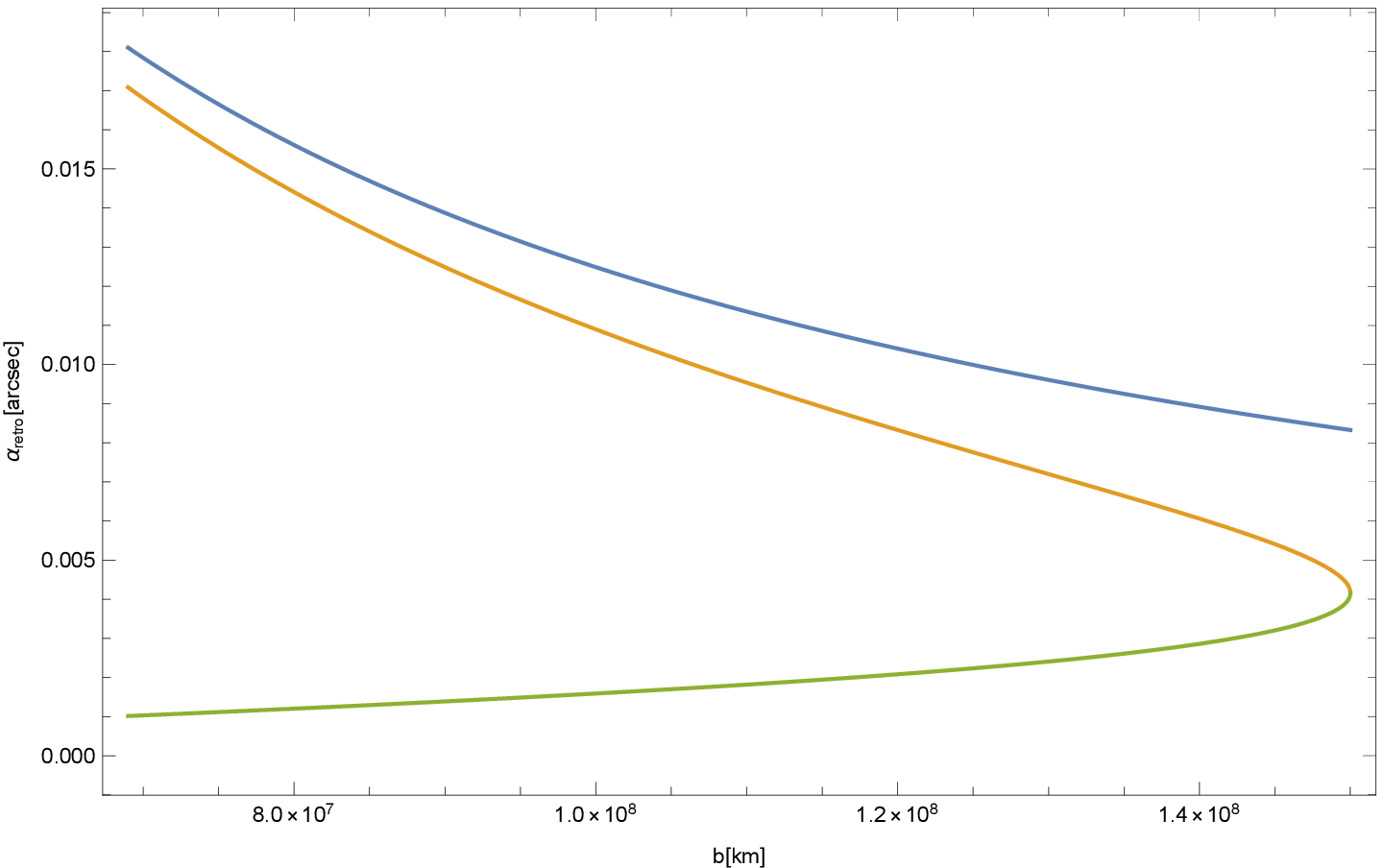}
\caption{ 
$\alpha$ for light of retrograde motion.
The horizontal axis denotes the impact parameter for a photon orbit 
and the vertical axis denotes the deflection angle of light.
The blue curve is the asymptotic deflection angle by the Kerr black hole.
The orange curve is the deflection angle with finite-correction by 
the Kerr black hole.
The green curve shows the difference between the asymptotic bending angle 
and the deflection angle with finite-correction by the Kerr black hole.}
\label{retdifference}
\end{figure}

See Figures \ref{difference} and \ref{retdifference} 
for the deflection angle with finite-distance corrections for 
the prograde motion and retrograde one, respectively, 
where we choose $r_S \sim 1.5\times 10^{8}$ km and $r_R \sim \infty$. 
The finite-distance correction reduces 
the deflection angle of light.
As the impact parameter $b$ increases, 
the finite-distance correction also increases.

As a second example, 
we discuss Sgr A$^{\ast}$ that is located 
at our galactic center. 
This object is a good candidate 
for measuring the strong gravitational 
deflection of light. 
The distance to the receiver is much larger 
than the impact parameter of light. 
On the other hand, some of source stars may live 
in our galactic center. 

For Sgr A$^{\ast}$, Eq. (\ref{delta-alpha2-axial}) 
becomes 
\begin{align}
\left| \delta\alpha_{GM} \right| 
&\sim 
s \left( \frac{M}{r_S} \right)^2 
\nonumber\\
&\sim 
10^{-7} \mbox{arcsec.} 
\times 
\left(\frac{s}{0.1}\right) 
\left(\frac{M}{4 \times 10^6 M_{\odot}}\right)^2 
%\times 
\left(\frac{0.1 \mbox{pc}}{r_S}\right)^2 , 
\label{alpha-Sgr-axial}
\end{align}
where we assume that the mass of the central black hole is $M \sim 4 \times 10^6 M_{\odot}$. 
This correction is nearly at a sub-microarcsecond level. Therefore, it is 
beyond the capability of present technology (e.g. \cite{EHT}). 

See Figure \ref{fig-Sgr} for 
the finite-distance correction due to the source location. 
The result in this figure is in agreement with 
the above order-of-magnitude estimation. 
This figure suggests 
the very weak dependence on the impact parameter $b$.

%\begin{figure}%[H]
%\includegraphics[width=10cm]{fig-4.eps}
%\caption{ 
%$\delta\alpha_{GM}$ for Sgr A$^{\ast}$. 
%The vertical axis denotes the finite-distance correction 
%to the deflection angle of light, 
%while the horizontal axis denotes the source distance $r_S$. 
%The solid curve (blue in color) and dashed one (red in color) 
%correspond to $b=10^2 M$ and $b=10^4 M$, respectively. 
%The dotted line (yellow in color) denotes the leading term 
%in $\delta\alpha_{GM}$ given by Eq. (\ref{delta-alpha}). 
%The overlap between these plots means that 
%$\delta\alpha_{GM}$ depends faintly on the impact parameter. $b$. 
%}
%\label{fig-Sgr}
%\end{figure}

\section{Rotating Teo wormhole: Another example} 
\subsection{Rotating Teo wormhole and optical metric}
In this section, we consider a rotating Teo wormhole \cite{Teo} 
in order to examine how our method can be applied 
to a wormhole spacetime. 
The spacetime metric for this wormhole is 
\begin{align}
ds^2=&-N^2dt^2+\frac{dr^2}{1-\frac{b_0}{r}}
+r^2H^2\Big[d\theta^2+\sin^2\theta(d\phi-\omega dt)^2 \Big] ,
\label{ds2RTWH}
\end{align}
where we denote 
\begin{align}
N=&H=1+\frac{d(4 \bar{a} \cos\theta)^2}{r}, 
\label{N}
\\
\omega=&\frac{2 \bar{a}}{r^3} . 
\label{omega}
\end{align}
Here, 
$b_0$ means the throat radius of this wormhole, 
$\bar{a}$ is corresponding to the spin angular momentum, 
and $d$ is a positive constant. 

For the rotating Teo wormhole Eq.(\ref{ds2RTWH}), 
the components of the generalized optical metric are \cite{Ono2018} 
\begin{align}
\gamma_{ij}dx^idx^j=&
\frac{r^7}{(r-b_0) \left(r^4-4\bar{a}^2\sin^2\theta \right) 
\left(16d\bar{a}^2\cos^2\theta +r\right)^2}dr^2 \notag\\
&+\frac{r^6}{r^4-4\bar{a}^2\sin^2\theta } d\theta^2
+\frac{r^{10}\sin^2\theta }{\left(r^4-4\bar{a}^2\sin^2\theta \right)^2}d\phi^2 . 
\label{gamma_RTWH}
\end{align}
Here, $\gamma_{ij}$ is not the induced metric in the ADM formulation. 
The components of $\beta_i$ are obtained as 
\begin{align}
\beta_idx^i=&-\frac{2 \bar{a} r^3 \sin^2\theta }{r^4-4 \bar{a}^2 \sin^2\theta } 
d\phi .
\label{beta_RTWH}
\end{align}

In this section, we restrict ourselves 
within the equatorial plane, 
namely $\theta = \pi/2$. 
On the equatorial plane, 
the constant $d$ in the metric always vanish, 
because $d$ is always associated 
with $\cos\theta$.  

We employ the same way for the Kerr case, 
we first derive the orbit equation 
on the equatorial plane from Eq.(\ref{OE}) as
\begin{align}
\left(\frac{dr}{d\phi}\right)^2=&
-\frac{r^5 (b_0-r) \left(4 \bar{a}^2 b^2-4 \bar{a} b r^3-b^2 r^4+r^6\right)}
{\left(-4 \bar{a}^2 b+2 \bar{a} r^3+b r^4\right)^2} \notag\\
=&\frac{r^4}{b^2}-r^2-\frac{{b_0}r^3}{b^2}+{b_0}r-\frac{4\bar{a}r^3}{b^3}
+\frac{4\bar{a}b_0r^2}{b^3}
+\mathcal{O}(\bar{a}^2/b^2) ,
\label{drdphiRTWH}
\end{align}
where $b$ denotes the impact parameter of 
the light ray and 
we use the weak field and slow rotation approximations  in the last line. 
There are no $b_0$ squared terms in the last line.
The orbit equation thus becomes 
\begin{align}
\left(\frac{du}{d\phi}\right)^2 
=&\frac{1}{b^2}-u^2-\frac{b_0u}{b^2}+b_0u^3-\frac{4\bar{a}u}{b^3}
-\frac{4\bar{a}b_0u^2}{b^3}
+\mathcal{O}(\bar{a}^2/b^6) . 
\label{dudphiRTWH}
\end{align} 
This equation is iteratively solved as 
\begin{align}
u=\frac{\sin\phi}{b}+\frac{\cos^2\phi}{2b^2}b_0-\frac{2}{b^3}\bar{a}
+\mathcal{O}\left(\frac{{b_0}^2}{b^3}, \frac{\bar{a}b_0}{b^4}\right) .
\label{uRTWH}
\end{align}
Solving Eq.(\ref{uRTWH}) for $\phi_S$ and $\phi_R$, 
we obtain $\phi_S$ and $\phi_R$ as
\begin{align}
\phi_S=&\arcsin(bu_S)-\frac{b_0\sqrt{1-b^2{u_S}^2}}{2b}
+\frac{2\bar{a}}{b^2\sqrt{1-b^2{u_S}^2}}
+\mathcal{O}\left(\frac{{b_0}^2}{b^2}, \frac{\bar{a}b_0}{b^3}\right) , 
\label{phiSRTWH} \\
\phi_R=&\pi-\arcsin(bu_R)+\frac{b_0\sqrt{1-b^2{u_R}^2}}{2b}
-\frac{2\bar{a}}{b^2\sqrt{1-b^2{u_R}^2}}
+\mathcal{O}\left(\frac{{b_0}^2}{b^2}, \frac{\bar{a}b_0}{b^3}\right) . 
\label{phiRRTWH}
\end{align}

\subsection{Gaussian curvature}
In the weak field approximation, 
the Gaussian curvature of the equatorial plane 
is 
\begin{align}
K
%=&\frac{R_{r\phi r\phi}}{\det\gamma}
%\notag\\
%=&\frac{1}{\sqrt{\det\gamma}}\Big[\frac{\partial}{\partial\phi}
%\Big(\frac{\sqrt{\det\gamma}}{\gamma_{rr}}{}^{(3)}\Gamma^{\phi}_{~rr}\Big)
%-\frac{\partial}{\partial r}
%\Big(\frac{\sqrt{\det\gamma}}{\gamma_{rr}{}^{(3)}}\Gamma^{\phi}_{~r\phi}\Big)
%\Big] 
%\notag\\
=&-\frac{b_0}{2 r^3}-\frac{56 \bar{a}^2}{r^6}
+ \mathcal{O}\left(\frac{\bar{a}^2b_0}{r^7}, \frac{\bar{a}^4}{r^{10}}\right) , 
\end{align}
where 
%$\gamma$ denotes $\det(\gamma_{ij})$, and 
$\bar{a}$ and $b_0$ play a role of 
book-keeping parameters 
in the weak field approximation. 
It is not surprising that this Gaussian curvature 
deviates from Eq. (26) in Jusufi and \"Ovg\"un \cite{JO}, 
because their Gaussian curvature describes 
a different surface that is defined with 
using the Randers-Finsler metric. 
The Randers-Finsler metric is quite 
different from our generalized optical metric $\gamma_{ij}$. 

When we perform the surface integral of the Gaussian curvature in Eq. (\ref{GB-axial}), 
we use Eq.(\ref{uRTWH}) for a boundary 
of the integration domain.  
The surface integral of the Gaussian curvature in Eq. (\ref{GB-axial}) 
is thus calculated as 
\begin{align}
-\iint_{{}^{\infty}_{R}{\square}^{\infty}_{S}} K dS 
=&
\int_{\phi_S}^{\phi_R}\int_{\infty}^{r(\phi)} \left(-\frac{b_0}{2r^2}\right) 
dr d\phi 
+\mathcal{O}\left(\frac{{b_0}^2}{b^2}, \frac{\bar{a}b_0}{b^3}\right)
\notag\\
=&\frac{b_0}{2}
\int_{\phi_S}^{\phi_R} 
\int_{0}^{\frac{\sin\phi}{b}+\frac{\cos^2\phi}{2b^2}b_0-\frac{2}{b^3}\bar{a}} dud\phi  
+\mathcal{O}\left(\frac{{b_0}^2}{b^2}, \frac{\bar{a}b_0}{b^3}\right)
\notag\\
%=&\frac{b_0}{2}\int_{\phi_S}^{\phi_R}
%\Big[u\Big]^{\frac{\sin\phi}{b}+\frac{\cos^2\phi}{2b^2}b_0-\frac{2}{b^3}a}_{0} d\phi 
%+\mathcal{O}({b_0}^2,ab_0)
%\notag\\
=&\frac{b_0}{2}\int_{\phi_S}^{\phi_R}
\Big[\frac{\sin\phi}{b}\Big] d\phi 
+\mathcal{O}\left(\frac{{b_0}^2}{b^2}, \frac{\bar{a}b_0}{b^3}\right)
\notag\\
=&\frac{b_0}{2}\Big[-\frac{\cos\phi}{b}\Big]_{\phi=\phi_S}^{\phi_R} 
+\mathcal{O}\left(\frac{{b_0}^2}{b^2}, \frac{\bar{a}b_0}{b^3}\right)
\notag\\
=&\frac{b_0}{2b}
\left(\sqrt{1-b^2{u_R}^2}+\sqrt{1-b^2{u_S}^2}\right)
+\mathcal{O}\left(\frac{{b_0}^2}{b^2}, \frac{\bar{a}b_0}{b^3}\right) ,
\label{intK}
\end{align}
where we use $\sin{\phi_R} = bu_R +\mathcal{O}(\bar{a}b^{-2}, b_0b^{-1})$ 
and $\sin{\phi_S} = bu_S +\mathcal{O}(\bar{a}b^{-2}, b_0b^{-1})$ 
by Eqs.(\ref{phiRRTWH}) and (\ref{phiSRTWH}) in the last line.

\subsection{Geodesic curvature of photon orbit}
We study 
the geodesic curvature of the photon orbit 
on the equatorial plane in the stationary and axisymmetric spacetime 
by using the generalized optical metric. 
It generally becomes  
\cite{Ono2017}
\begin{align}
\kappa_g=-\sqrt{\frac{1}{\gamma\gamma^{\theta\theta}}}\beta_{\phi,r} . 
\label{kappa_equ}
\end{align}
In the Teo wormhole, 
this expression is rearranged as 
\begin{align}
\kappa_g=&-\frac{2\bar{a}}{r^3}+\frac{\bar{a}b_0}{r^4}
+\frac{\bar{a}{b_0}^2}{4r^5}+\frac{\bar{a}{b_0}^3}{8r^6}\
+\mathcal{O}\left(\frac{\bar{a}^3}{r^7}, \frac{\bar{a}^3b_0}{r^8}\right) . 
\end{align}

We compute the path integral of the geodesic curvature  of the photon orbit. 
The detailed calculations and result are   
\begin{align}
\int_S^R\kappa_gd\ell
=&\int^{S}_{~R}\frac{2\bar{a}}{r^3} d\ell  
+\mathcal{O}\left(\frac{{b_0}^2}{b^2}, \frac{\bar{a}b_0}{b^3}\right)
\notag\\
=&\int^{\pi/2-\phi_S}_{~\pi/2-\phi_R} \frac{2\bar{a}\cos\vartheta}{b^2}
d\vartheta 
+\mathcal{O}\left(\frac{{b_0}^2}{b^2}, \frac{\bar{a}b_0}{b^3}\right)
\notag\\
=&\frac{2\bar{a}}{b^2}
\left[\sin\left(\frac{\pi}{2}-\phi_S\right)
-\sin\left(\frac{\pi}{2}-\phi_R\right)\right] 
+\mathcal{O}\left(\frac{{b_0}^2}{b^2}, \frac{\bar{a}b_0}{b^3}\right)
\notag\\
=&\frac{2\bar{a}}{b^2}\left(\sqrt{1-b^2{u_S}^2}+\sqrt{1-b^2{u_R}^2}\right) 
+\mathcal{O}\left(\frac{{b_0}^2}{b^2}, \frac{\bar{a}b_0}{b^3}\right) , 
\label{intkappa}
\end{align}
for the retrograde orbit of the photon. 
In the last line, we used 
$\sin{\phi_R} = bu_R +\mathcal{O}(\bar{a}b^{-2}, b_0b^{-1})$ 
and $\sin{\phi_S} = bu_S +\mathcal{O}(\bar{a}b^{-2}, b_0b^{-1})$ 
from Eq. (\ref{uRTWH}). 
The above result becomes $4\bar{a}/b^2$, as $r_R \to \infty$ and 
$r_S \to \infty$. 
The sign of the right hand side 
in Eq. (\ref{intkappa}) is opposite, 
if the photon is in the prograde motion.

\subsection{$\phi_{RS}$ part}
The rotating Teo wormhole is an asymptotically flat spacetime 
as seen from Eq.(\ref{ds2RTWH}).
Therefore, 
the integral of the geodesic curvature of the circular arc segment with an infinite radius 
can be expressed simply as $\phi_{RS}$. 
By using Eqs.(\ref{phiSRTWH}) and (\ref{phiRRTWH}), 
$\phi_{RS}$ is obtained as
\begin{align}
\phi_{RS}=&\phi_R-\phi_S \notag\\
=&\pi-\arcsin(bu_R)-\arcsin(bu_S)
+\frac{b_0\sqrt{1-b^2{u_R}^2}}{2b}+\frac{b_0\sqrt{1-b^2{u_S}^2}}{2b}\notag\\
&-\frac{2\bar{a}}{b^2\sqrt{1-b^2{u_R}^2}}
-\frac{2\bar{a}}{b^2\sqrt{1-b^2{u_S}^2}}
+\mathcal{O}\left(\frac{{b_0}^2}{b^2}, \frac{\bar{a}b_0}{b^3}\right) . 
\label{phiRSRTWH}
\end{align} 

\subsection{$\Psi$ parts}
For the rotating Teo wormhole by Eq.(\ref{ds2RTWH}), 
Eq.(\ref{sinPsiR}) is computed as 
\begin{align}
\sin\Psi_R=&bu_R+2\bar{a}{u_R}^2-4\bar{a}^2b{u_R}^5, \label{sinPsiRRTWH}
\end{align}
and Eq.(\ref{sinPsiS}) becomes
\begin{align}
\sin(\pi-\Psi_S)=&bu_S+2\bar{a}{u_S}^2-4\bar{a}^2b{u_S}^5, \label{sinPsiSRTWH}
\end{align}
where the slow rotation approximation is not needed. 
Therefore, we obtain $\Psi_R$ and $\Psi_S$ as
\begin{align}
\Psi_R=&\arcsin(b {u_R})+\frac{2 \bar{a} {u_R}^2}{\sqrt{1-b^2 {u_R}^2}}
+\frac{2 \bar{a}^2 b {u_R}^5 \left(2 b^2 {u_R}^2-1\right)}
{\left(b^2 {u_R}^2-1\right)^{3/2}}
+\mathcal{O}(\bar{a}^3/b^6), \label{PsiRRTWH} \\
\pi-\Psi_S=&\arcsin(b {u_S})+\frac{2 \bar{a} {u_S}^2}{\sqrt{1-b^2 {u_S}^2}}
+\frac{2 \bar{a}^2 b {u_S}^5 \left(2 b^2 {u_S}^2-1\right)}
{\left(b^2 {u_S}^2-1\right)^{3/2}}+\mathcal{O}(\bar{a}^3/b^6), \label{PsiSRTWH}
\end{align}
where we used the slow rotation approximation.

\subsection{Deflection angle of light}
We combine Eqs. (\ref{intK}) and (\ref{intkappa}) 
to obtain 
the deflection angle of light in the prograde orbit
as 
\begin{align}
\alpha_{\mbox{prog}}
=&\frac{b_0}{2b}
\left(\sqrt{1-b^2{u_R}^2}+\sqrt{1-b^2{u_S}^2}\right)
-\frac{2\bar{a}}{b^2}
\left(\sqrt{1-b^2{u_R}^2}+\sqrt{1-b^2{u_S}^2}\right) \notag\\
&+\mathcal{O}\left(\frac{{b_0}^2}{b^2}, \frac{\bar{a}b_0}{b^3}\right) . 
\label{alpha-prog}
\end{align}
The deflection angle of the retrograde light is 
\begin{align}
\alpha_{\mbox{retro}}=&
\frac{b_0}{2b}
\left(\sqrt{1-b^2{u_R}^2}+\sqrt{1-b^2{u_S}^2}\right)
+\frac{2\bar{a}}{b^2}
\left(\sqrt{1-b^2{u_R}^2}+\sqrt{1-b^2{u_S}^2}\right) \notag\\
&+\mathcal{O}\left(\frac{{b_0}^2}{b^2}, \frac{\bar{a}b_0}{b^3}\right) . 
\label{alpha-retro}
\end{align}

Next, by using Eqs. (\ref{phiRSRTWH}), (\ref{PsiRRTWH}) and (\ref{PsiSRTWH}), 
we obtain the deflection angle 
of the prograde light as 
\begin{align}
\alpha_{\mbox{prog}}
=&\pi-\arcsin(bu_R)-\arcsin(bu_S)
+\frac{b_0\sqrt{1-b^2{u_R}^2}}{2b}+\frac{b_0\sqrt{1-b^2{u_S}^2}}{2b}\notag\\
&-\frac{2\bar{a}}{b^2\sqrt{1-b^2{u_R}^2}}-\frac{2\bar{a}}{b^2\sqrt{1-b^2{u_S}^2}}
+\arcsin(b {u_R})+\frac{2 \bar{a} {u_R}^2}{\sqrt{1-b^2 {u_R}^2}} \notag\\
&-\pi+\arcsin(b {u_S})+\frac{2 \bar{a} {u_S}^2}{\sqrt{1-b^2 {u_S}^2}} 
+\mathcal{O}\left(\frac{{b_0}^2}{b^2}, \frac{\bar{a}b_0}{b^3}\right) \notag\\
=& \frac{b_0}{2b}
\left(\sqrt{1-b^2{u_R}^2}+\sqrt{1-b^2{u_S}^2}\right)
-\frac{2\bar{a}}{b^2}
\left(\sqrt{1-b^2{u_R}^2}+\sqrt{1-b^2{u_S}^2}\right) \notag\\
&+\mathcal{O}\left(\frac{{b_0}^2}{b^2}, \frac{\bar{a}b_0}{b^3}\right) . 
\label{Alpha-prog}
\end{align}
The deflection angle of light 
in the retrograde orbit is 
\begin{align}
\alpha_{\mbox{retro}}=&
\frac{b_0}{2b}
\left(\sqrt{1-b^2{u_R}^2}+\sqrt{1-b^2{u_S}^2}\right)
+\frac{2\bar{a}}{b^2}
\left(\sqrt{1-b^2{u_R}^2}+\sqrt{1-b^2{u_S}^2}\right) \notag\\
&+\mathcal{O}\left(\frac{{b_0}^2}{b^2}, \frac{\bar{a}b_0}{b^3}\right) . 
\label{Alpha-retro}
\end{align}

The deflection of light in the prograde (retrograde) orbit is weaker 
(stronger) with increasing the angular momentum of the Teo wormhole. 
The reason is as follows. 
The local inertial frame 
in which the light travels at the light speed $c$ 
in general relativity moves faster (slower). 
Hence, the time-of-flight of light becomes 
shorter (longer). 
On the light propagation 
A similar explanation is done by 
using the dragging of the inertial frame 
also by Laguna and Wolsczan \cite{LW}. 
They discussed the Shapiro time delay. 
The expression of the deflection angle of light 
by a rotating Teo wormhole is similar 
to that by Kerr black hole. 
This implies that it is hard to distinguish 
Kerr black hole from rotating Teo wormhole 
by the gravitational lens observations. 

In Eqs.(\ref{Alpha-prog}) and (\ref{Alpha-retro}), 
the source and receiver can be located at finite  distance from the wormhole. 
In the limit as $r_R \to \infty$ and $r_S \to \infty$, 
Eqs. (\ref{alpha-prog}) and (\ref{alpha-retro}) become 
\begin{align}
\alpha_{\mbox{prog}} \rightarrow \frac{b_0}{b}-\frac{4\bar{a}}{b^2} 
+\mathcal{O}\left(\frac{{b_0}^2}{b^2}, \frac{\bar{a}b_0}{b^3}\right) ,
\notag\\
\alpha_{\mbox{retro}} \rightarrow \frac{b_0}{b}+\frac{4\bar{a}}{b^2} 
+\mathcal{O}\left(\frac{{b_0}^2}{b^2}, \frac{\bar{a}b_0}{b^3}\right) .
\end{align}
They are in complete agreement 
with Eqs. (39) and (56) in Jusufi and \"Ovg\"un \cite{JO}, 
where they restrict themselves within the asymptotic source and receiver 
($r_R \to \infty$ and $r_S \to \infty$).

\subsection{Finite-distance corrections 
in the Teo wormhole spacetime}
To be precise, we define 
the finite-distance correction to 
the deflection angle of light 
as the difference between 
the asymptotic deflection angle $\alpha_{\infty}$ 
and the deflection angle for the finite distance case. 
It is denoted as $\delta \alpha$. 

\begin{figure}
\includegraphics[width=15cm]{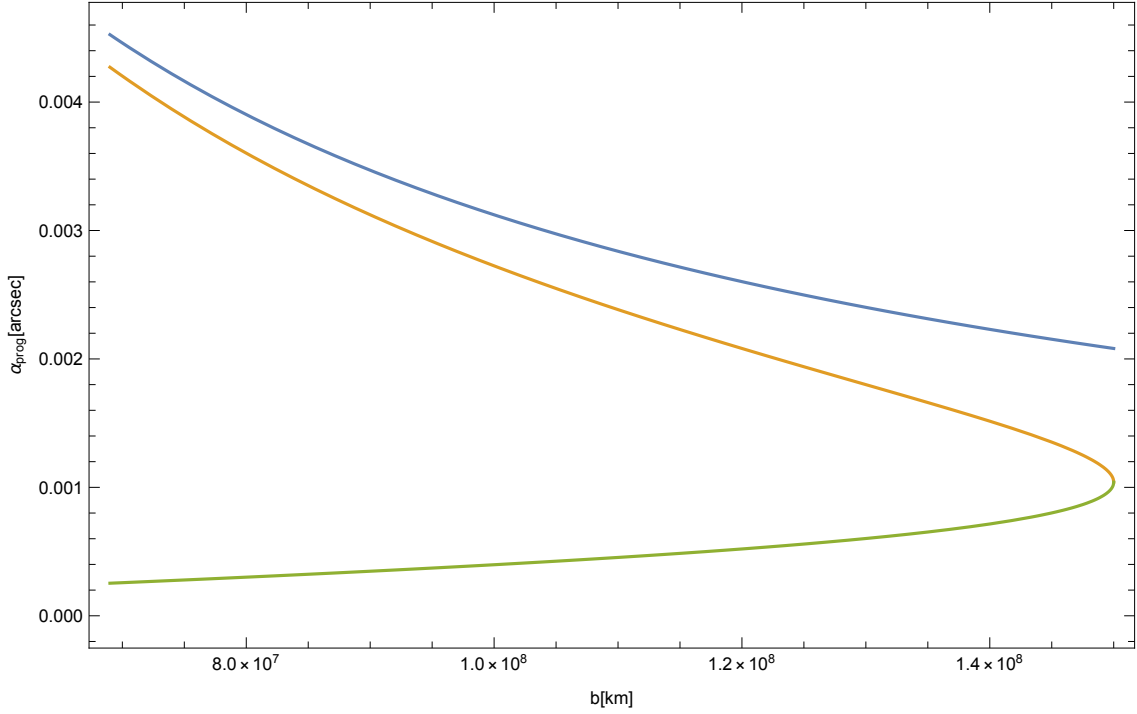}
\caption{ 
$\alpha$ in the Teo wormhole. 
The blue curve is the asymptotic deflection angle 
by the rotating Teo wormhole. 
The orange curve is the deflection angle with finite-distance corrections 
by the rotating Teo wormhole.
The blue curve shows the difference between the asymptotic deflection angle and 
the deflection angle with finite-distance corrections 
by the rotating Teo wormhole.}
\label{RTWHdiff}
\end{figure}

We consider the following situation. 
An observer on the Earth 
sees the light deflected by the solar mass. 
The source o light is located 
in a practically asymptotic region.
In other words, we choose $ b_0=M_{\odot}$, $\bar{a}=J_{\odot}$, 
$r_R \sim 1.5\times 10^{8}$ km, $r_S \sim \infty$. 
See Figure \ref{RTWHdiff} for 
the finite-distance correction due to the impact parameter $b$. 
In Figure \ref{RTWHdiff}, the green curve means the difference between 
the asymptotic bending angle and the deflection angle with finite-distance corrections, 
the blue curve denotes the asymptotic deflection angle and 
the orange curve is the deflection angle with finite-distance corrections 
by the rotating Teo wormhole.  
The deflection angle is decreased by 
the finite-distance correction.  
If the impact parameter $b$ increases, 
the finite-distance correction also increases.

\begin{figure}
\includegraphics[width=15cm]{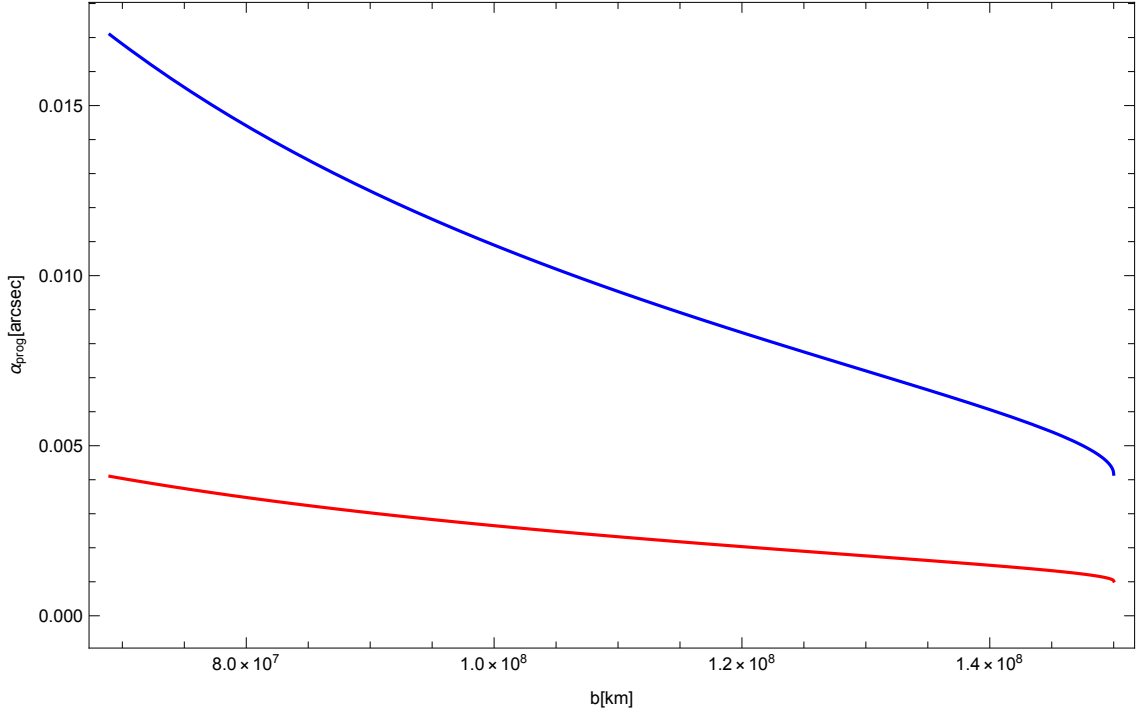}
\caption{ 
$\alpha$ for prograde motion of light. 
The horizontal axis is the impact parameter of photon orbit. 
The vertical axis means the deflection angle 
of light.  
The blue curve means the deflection angle with 
finite-distance corrections by the Kerr black hole. 
The red curve corresponds to that by the rotating Teo wormhole.
For the purpose of this comparison, 
the mass of a Kerr black hole $M$ and the throat radius of a 
rotating Teo wormhole $b_0$ are chosen as $M=b_0=M_{\odot}$. 
The spin angular momentum of a Kerr black hole and that of a rotating Teo wormhole 
are chosen as the same as that the Sun for the simplicity. 
}
\label{RTWHKerr}
\end{figure}

See also Figure \ref{RTWHKerr} for numerical calculations of 
the finite-distance correction due to the impact parameter $b$. 
In Figure \ref{RTWHKerr},  
the blue curve is the deflection angle with finite-distance correction 
by a Kerr black hole and 
the red curve is the deflection angle with finite-correction by a rotating Teo wormhole.  
The deflection of light is stronger 
in a Kerr black hole case 
for the chosen values.

\section{Summary}
In this paper, we provided a brief review of 
a series of works on the deflection angle of light 
for 
a light source and receiver 
in a non-asymptotic region. 
\cite{Ishihara2016,Ishihara2017,Ono2017,Ono2018}. 
The validity and usefulness of the new formulation 
come from the GB theorem in differential geometry.  
First, we discussed how to define 
the gravitational deflection angle of light 
in a static, spherically symmetric and asymptotically flat spacetime, 
for which we assume 
the finite-distance source and receiver. 
We examined whether our definition 
is invariant geometrically by using the GB theorem. 
By using our definition, we carefully 
computed finite-distance corrections 
to the light deflection in Schwarzschild spacetime. 
We considered 
both cases of weak deflection and 
strong one. 
Next, we extended the definition to stationary and axisymmetric spacetimes. 
This extension allows us to compute finite-distance corrections for Kerr black holes and rotating Teo wormholes. 
We verified that these results are consistent with the previous works in the infinite-distance limit.
We mentioned also the finite-distance corrections 
to the light 
deflection by Sagittarius A$^*$. 
It is left as future work to apply the present formulation 
to other interesting spacetime models 
and also to extend it
to a more general spacetime structure.

%\begin{acknowledgments} 
\acknowledgments{
We are grateful to Marcus Werner for the stimulating and very fruitful discussions. 
We thank Takao Kitamura, Asahi Ishihara and Yusuke Suzuki 
for useful conversations. 
We would like to thank 
Yuuiti Sendouda, Ryuichi Takahashi, Yuya Nakamura and 
Naoki Tsukamoto 
for the useful conversations. 
This work was supported 
in part by JSPS research fellowship for young researchers (T.O.),  
in part by Japan Society for the Promotion of Science 
Grant-in-Aid for Scientific Research, 
No. 18J14865 (T.O.), No. 17K05431 (H.A.), 
and in part by by Ministry of Education, Culture, Sports, Science, and Technology,  
No. 17H06359 (H.A.).
}
%\end{acknowledgments}

\appendix
\section{Detailed calculations at $O(M^2/b^2)$ and $O(a^2/b^2)$ in Kerr spacetime}
First, we investigate the Gaussian curvature $K$ of the equatorial plane 
in the Kerr spacetime. 
Here, we assume the weak field and slow rotation approximations. 
Up to the second order, $K$ is expanded as 
\begin{align}
K&=\frac{R_{r\phi r\phi}}{\gamma} 
\notag\\
&= -\frac{2M}{r^3}+\frac{3M^2}{r^4}
+O\left(\frac{a^2M}{r^5}\right) , 
\label{K-second}
\end{align}
where $\gamma$ denotes $\det{(\gamma_{ij})}$. 
There are no $a^2$ terms in $K$. 
More interestingly, only the $a^2M$ term at the third order level 
do exist in $K$. 
By noting that $K$ begins with $O(M)$, what we need 
for the second-order calculations is 
only the linear-order term in the area element on the equatorial plane. 
This is obtained as 
\begin{align}
dS \equiv& \sqrt{\gamma} drd\phi \notag\\
=& \left[r + 3M + O\left(\frac{M^2}{r}\right) \right] drd\phi , 
\label{dS-second} 
\end{align}
where terms at $O(a)$ and also at $O(a^2)$ do not exist in $dS$. 
This is because all terms including the spin parameter 
cancel out in $\gamma$ for $\theta = \pi/2$ 
and $\gamma$ thus depends only on $M$,  
as shown by direct calculations. 
%It is rather clear that no terms appear at $O(a)$ 
%in the right-hand side of Eq. (\ref{dS-second}). 
%Moreover, by direct caluculations, 
%one can see that terms at $O(a^2)$ cencel out 
%in the right-hand side of Eq. (\ref{dS-second}). 

By using Eqs. (\ref{K-second}) and (\ref{dS-second}), 
the surface integration of the Gaussian curvature 
is performed as  
\begin{align}
-\iint K dS 
=&\int_{\infty}^{r_{OE}} dr
\int_{\phi_S}^{\phi_R} d\phi \Big(-\frac{2M}{r^3}+\frac{3M^2}{r^4}\Big)
(r+3M) 
+ O\left(\frac{M^3}{b^3}, \frac{aM^2}{b^3}, \frac{a^2M}{b^3}\right)
\notag\\
=&\int_{0}^{\frac{1}{b}\sin\phi+\frac{M}{b^2}(1+\cos^2\phi)} du
\int_{\phi_S}^{\phi_R} d\phi ~(2M+3uM^2) 
+ O\left(\frac{M^3}{b^3}, \frac{aM^2}{b^3}, \frac{a^2M}{b^3}\right)
\notag\\
%=&\int_{\phi_S}^{\phi_R}
%\Big[2uM+\frac{3u^2}{2}M^2\Big]^{\frac{1}{b}\sin\phi+\frac{M%}{b^2}(1+\cos^2\phi)}_{0}
% d\phi 
%+ O\left(\frac{M^3}{b^3}, \frac{aM^2}{b^3}, %\frac{a^2M}{b^3}\right)
%\notag\\
=&\int_{\phi_S}^{\phi_R}
\Big[\frac{2M}{b}\sin\phi+\frac{M^2}{2b^2}(7+\cos^2\phi)\Big] d\phi
+ O\left(\frac{M^3}{b^3}, \frac{aM^2}{b^3}, \frac{a^2M}{b^3}\right)
\notag\\
=&\frac{2M}{b}\Big[\cos\phi\Big]^{\phi_S}_{\phi_R}
+\frac{M^2}{2b^2} \Big[\frac{30\phi+\sin(2\phi)}{4}\Big]^{\phi_R}_{\phi_S}
+ O\left(\frac{M^3}{b^3}, \frac{aM^2}{b^3}, \frac{a^2M}{b^3}\right)
\notag\\
%=&\frac{2M}{b}\Big[\sqrt{1-b^2{u_S}^2}+\sqrt{1-b^2{u_R}^2} %\Big] 
%\notag\\
%&+\frac{2M^2}{b}\Big[\frac{u_S(2-b^2{u_S}^2)}{\sqrt{1-b^2{u_%S}^2}}
%+\frac{u_R(2-b^2{u_R}^2)}{\sqrt{1-b^2{u_R}^2}}\Big]
%\notag\\
%&+\frac{15 M^2}{4b^2}[\pi-\arcsin(bu_S)-\arcsin(bu_R)]
%\notag\\
%&-\frac{M^2}{4b^2}[bu_S\sqrt{1-b^2{u_S}^2}+bu_R\sqrt{1-b^2{u%_R}^2}] 
%+ O\left(\frac{M^3}{b^3}, \frac{aM^2}{b^3}, %\frac{a^2M}{b^3}\right)
%\notag\\
=&\frac{2M}{b}\Big[\sqrt{1-b^2{u_S}^2}+\sqrt{1-b^2{u_R}^2}\Big] 
\notag\\
&+\frac{15 M^2}{4b^2}[\pi-\arcsin(bu_S)-\arcsin(bu_R)]
\notag\\
&+\frac{M^2}{4b^2}[\frac{bu_S(15-7b^2{u_S}^2)}{\sqrt{1-b^2{u_S}^2}}
+\frac{bu_R(15-7b^2{u_R}^2)}{\sqrt{1-b^2{u_R}^2}}] 
+ O\left(\frac{M^3}{b^3}, \frac{aM^2}{b^3}, \frac{a^2M}{b^3}\right) , 
\label{int-K-second}
\end{align}
where we use, in the second line, 
an iterative solution for the orbit equation 
by Eq. (\ref{OE}) in the Kerr spacetime. 

Next, we study the geodesic curvature. 
On the equatorial plane, we find  
\begin{align}
\kappa_g=&-\frac{1}{\sqrt{\cfrac{\Sigma^2}{\Delta(\Sigma-2Mr)}
\left(r^2+a^2+\cfrac{2a^2Mr\sin^2\theta}{\Sigma}\right)
\cfrac{\Sigma\sin^2\theta}{(\Sigma-2Mr)}}}\beta_{\phi,r} \notag \\
=&-\frac{2aM}{r^3} +O\left( \frac{aM^2}{r^3}\right) . 
\label{kappag-second}
\end{align}
Note that $a^2$ terms do not exist. 
Therefore, we obtain 
\begin{align}
\int_{c_p}\kappa_gd\ell
=&-\int^{R}_{S} d\ell \left[ \frac{2aM}{r^2} 
+ O\left( \frac{aM^2}{r^3} \right) \right] \notag\\
=&-\frac{2aM}{b^2}\int^{\phi_R}_{~\phi_S} \cos\vartheta d\vartheta 
+ O\left( \frac{aM^2}{b^3} \right)
\notag\\
%=&-\frac{2aM}{b^2}[\sin\phi_R-\sin\phi_S] 
%+ O\left( \frac{aM^2}{b^3} \right)
%\notag\\
=&\frac{2aM}{b^2}[\sqrt{1-b^2{u_R}^2}+\sqrt{1-b^2{u_S}^2}] 
+ O\left( \frac{aM^2}{b^3} \right) , 
\label{int-kappag-second}
\end{align}
where we use 
$\sin\phi_S=\sqrt{{r_S}^2-b^2}/r_S + O(M/r_S)$ 
and  
$\sin\phi_R=-\sqrt{{r_R}^2-b^2}/r_R + O(M/r_R)$. 

By combining Eqs. (\ref{int-K-second}) and (\ref{int-kappag-second}), 
we obtain 
\begin{align}
\alpha \equiv& -\iint_{{}^{\infty}_{R}\square^{\infty}_{S}} K dS 
-\int^{S}_{R}\kappa_gd\ell \notag\\
=&\frac{2M}{b}\Big[\sqrt{1-b^2{u_S}^2}+\sqrt{1-b^2{u_R}^2}\Big] 
\notag\\
&+\frac{15 M^2}{4b^2}\left[\pi-\arcsin(bu_S)-\arcsin(bu_R)\right]
\notag\\
&+\frac{M^2}{4b^2}\left[\frac{bu_S(15-7b^2{u_S}^2)}{\sqrt{1-b^2{u_S}^2}}
+\frac{bu_R(15-7b^2{u_R}^2)}{\sqrt{1-b^2{u_R}^2}}\right] \notag\\
&-\frac{2aM}{b^2}\left[\sqrt{1-b^2{u_R}^2}+\sqrt{1-b^2{u_S}^2}\right] 
+ O\left(\frac{M^3}{b^3}, \frac{aM^2}{b^3}, \frac{a^2M}{b^3}\right) . 
\label{alpha-2nd}
\end{align}
Note that $a^2$ terms and $a^3$ ones do not appear in $\alpha$ 
for the finite distance situation as well as 
in the infinite distance limit. 
If we assume the infinite distance limit $u_R, u_S \to 0$, 
Eq. (\ref{alpha-2nd}) becomes 
\begin{align}
\alpha \rightarrow \frac{4M}{b}+\frac{15\pi M^2}{4b^2}-\frac{4aM}{b^2} . 
\end{align}
This agrees with the known results, 
especially on the numerical coefficients at the order of $M^2$ and $aM$.

%\newpage


\begin{thebibliography}{99}
\bibitem{Metric}
In this paper, we use the polar coordinates. 
In the cylindrical coordinates, the line element is known 
as the Weyl-Lewis-Papapetrou form \cite{Lewis,LR,Papapetrou}.
\bibitem{Kerr-bending}
Precise analytic treatments of the deflection angle of light 
were done in a conventional approach, 
on the equatorial plane of a Kerr black hole \cite{IH} 
and for generic photon orbits 
in terms of the generalized hypergeometric functions 
of Appell and Lauricella \cite{Kraniotis}. 
They assume that 
both the source and the receiver are located at the null infinity.
\end{thebibliography}
\end{document}